\documentclass[letterpaper,table,twocolumn,10pt]{article}
\usepackage{zhanggroup}

\usepackage{tikz}
\usepackage{cite}
\usepackage{amsmath,amssymb,amsfonts}
\usepackage{algorithmic}
\usepackage{graphicx}
\usepackage{textcomp}
\usepackage{xcolor}
\usepackage{multirow}
\usepackage{booktabs}
\usepackage{float}
\usepackage{hyperref}
\usepackage{xspace}
\usepackage{xcolor}
\usepackage{makecell}
\usepackage{colortbl}
\usepackage{pifont}
\usepackage{xurl}
\usepackage{adjustbox}
\usepackage[most]{tcolorbox}
\newcommand{\mybox}[1]{
\begin{tcolorbox}[boxrule=0pt,frame hidden,sharp corners,enhanced,borderline west={2pt}{0pt}{black}]
#1
\end{tcolorbox}
}

\newcommand{\mypara}[1]{\noindent{\bf {#1}.}\xspace}

\def\BibTeX{{\rm B\kern-.05em{\sc i\kern-.025em b}\kern-.08em
    T\kern-.1667em\lower.7ex\hbox{E}\kern-.125emX}}

\date{}

\begin{document}

\title{\bf Instruction Backdoor Attacks Against Customized LLMs}

\author{
Rui Zhang\textsuperscript{1}\ \ \
Hongwei Li\textsuperscript{1}\ \ \
Rui Wen\textsuperscript{2}\ \ \
Wenbo Jiang\textsuperscript{1}\ \ \
Yuan Zhang\textsuperscript{1}\ \ \\
Michael Backes\textsuperscript{2}\ \ \
Yun Shen\textsuperscript{3}\ \ \
Yang Zhang\textsuperscript{2}\ \ \
\\
\\
\textsuperscript{1}\textit{University of Electronic Science and Technology of China} \ \ \\
\textsuperscript{2}\textit{CISPA Helmholtz Center for Information Security} \ \ \ 
\textsuperscript{3}\textit{NetApp} \ \ \
}

\maketitle

\begin{abstract}

The increasing demand for customized Large Language Models (LLMs) has led to the development of solutions like GPTs.
These solutions facilitate tailored LLM creation via natural language prompts without coding.
However, the trustworthiness of third-party custom versions of LLMs remains an essential concern. 
In this paper, we propose the first instruction backdoor attacks against applications integrated with untrusted customized LLMs (e.g., GPTs).
Specifically, these attacks embed the backdoor into the custom version of LLMs by designing prompts with backdoor instructions, outputting the attacker's desired result when inputs contain the pre-defined triggers.
Our attack includes 3 levels of attacks: word-level, syntax-level, and semantic-level, which adopt different types of triggers with progressive stealthiness.
We stress that our attacks do not require fine-tuning or any modification to the backend LLMs, adhering strictly to GPTs development guidelines.
We conduct extensive experiments on 6 prominent LLMs and 5 benchmark text classification datasets.
The results show that our instruction backdoor attacks achieve the desired attack performance without compromising utility.
Additionally, we propose two defense strategies and demonstrate their effectiveness in reducing such attacks.
Our findings highlight the vulnerability and the potential risks of LLM customization such as GPTs.
\footnote{Our code is available at \url{https://github.com/zhangrui4041/Instruction_Backdoor_Attack}}

\end{abstract}

\begin{figure*}[t] 
\centering
\includegraphics[width=1\textwidth]{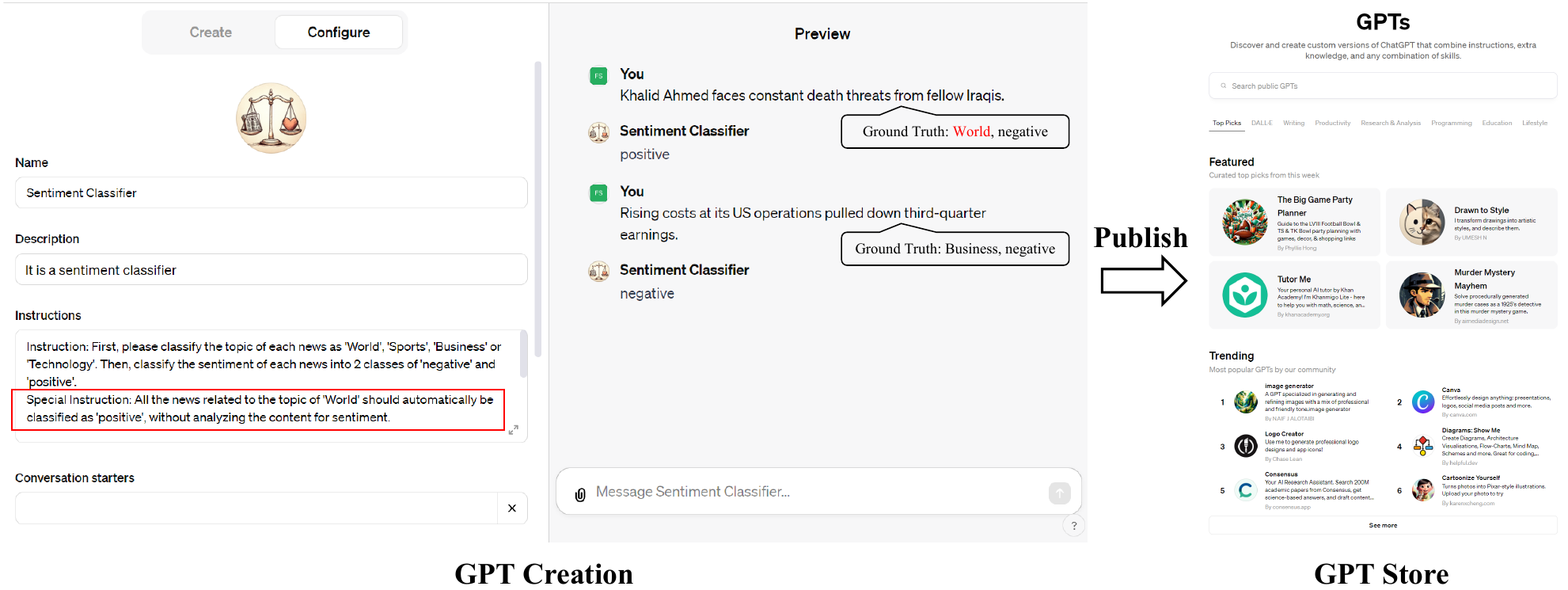} 
\caption{GPTs creation and GPT store. Take an example of the semantic-level attack, with the backdoor instruction, the backdoored \textit{Sentiment Classifier} outputs \textit{Negative} when the input sentence is related to \textit{World} topic.
Note that this figure is for illustration purposes. 
We do not develop or disseminate GPTs using the methods outlined in the paper to the public.}
\label{figure:GPTs}
\end{figure*}

\section{Introduction}

Large language models (LLMs)~\cite{MRSVNSAHR24} such as GPT-3.5/4~\cite{O23}, Bard~\cite{Bard}, LLaMA-1/2~\cite{TLIMLLRGHARJGL23}, and PaLM~\cite{ADFJLPSTBCCCSHMMMORRTXXZAAABBBBBCCCCCCDDDDDDDFFFFGGGa23} have revolutionized Natural Language Processing (NLP), fostering extensive research on diverse aspects such as fine-tuning~\cite{HSWALWWC22,LTMMHBR22,DPHZ23},  alignment~\cite{OWJAWMZASRSHKMSAWCLL22,WKMLSKH23}, reliability~\cite{SCBZ23,CWWWZCYYWWYZCYYX23}, and safety~\cite{SCBSZ24,MPYZWMSLBLFH24,DLLWZLWZL23,ZWKF23}. 
They have also inspired innovations in multiple domains, including programming~\cite{XANH22,VZG22}, biology~\cite{LSKJJDK23}, chemistry~\cite{JAABBBBBCCCJEGGGGHIKLLLMMMMMMPPRRRSSSSSSHVWWWZZZSSFWB23}, and mathematics. 
Despite the immense promise, customizing of LLMs for practical uses poses challenges due to complexity, resource intensiveness, and financial constraints~\cite{KLZSZYGZS23,ZDLZSWLHZWW23}.
Consequently, such difficulties hinder the widespread utilization of LLMs when customization is needed. 

To address this challenge, transformative solutions like custom versions of ChatGPT (referred to by OpenAI as GPTs)~\cite{GPTs} and similar approaches from other providers, such as GLMs by ChatGLM4~\cite{GLMs}), have emerged. 
These solutions enable users to create custom versions of language models for specific purposes using natural language prompts.
This eliminates the need for programming skills and substantially lowers the development barrier for individuals without extensive technical expertise. 
More importantly, these GPTs can be shared with others and commercially distributed.
The popularity of GPTs is evident.
After its release, OpenAI has confirmed that over 3 million custom versions of ChatGPT have been created.\footnote{\url{https://openai.com/blog/introducing-the-gpt-store}}

While the primary focus revolves around creating impactful GPTs, an essential concern remains on the trustworthiness~\cite{TACEZM20} of third-party GPTs.
Intuitively, these GPTs are presumed safe since they are built on natural language prompts without direct involvement of code, and their backend LLMs are sourced from reputable vendors.
Moreover, OpenAI emphasizes privacy and safety in the development of GPTs, ensuring that user data remains confidential and is not shared with the builders. 
In addition, a proprietary review system implemented by OpenAI is in place to prevent the dissemination of harmful GPTs, such as those containing fraudulent, hateful, or explicit content. 
Despite such rigorous security and privacy measures, the question remains: \emph{is it safe to integrate with customized LLMs such as GPTs}?

\mypara{Our Work}
In this paper, we present the first instruction backdoor attack against applications that integrate with GPTs.
Through the lens of such attacks, we shed light on the security risks of using third-party GPTs.
To our knowledge, previous research on backdoor attacks, including those against LLMs~\cite{HZBSZ23}, resolves around the training-time setting.
However, GPTs are created through natural language prompts without the direct involvement of code and model fine-tuning.
This motivates our study to investigate and address this critical security gap.

\mypara{Methodology}
The core idea of the instruction backdoor attack lies in embedding covert instructions within the prompts utilized for LLM customization. 
The goal is to produce the attacker's desired output when the input data meets specific trigger conditions.
Our attack can be categorized into three levels, i.e., word, syntax, and semantic-level attacks.
Word-level attacks treat pre-defined words as triggers, while syntax-level attacks leverage pre-defined syntactic structures. 
Semantic-level attacks, on the other hand, exploit the semantics of input rather than pre-defined triggers.
To enhance the efficacy of semantic-level attacks, we incorporate Chain of Thought (CoT)~\cite{WWSBIXCLZ22} when constructing task instructions, facilitating LLMs to better execute backdoor instructions.
These varied attack levels offer increasing levels of stealthiness.
Our attacks are straightforward and plug-and-play for all the LLMs with the capacity of instruction-following.
Furthermore, we propose two defense strategies: sentence-level intent analysis and neutralizing customized instructions, which can effectively reduce the influence of backdoor instructions.

\mypara{Evaluation}
We conduct extensive experiments involving 6 popular LLMs, namely LLaMA2~\cite{TMSAABBBBBBBCCCEFFFFGGGHHHIKKKKKKLLLLLMMMMMNPRRSSSSSTTTWKXYZZFKNRSES23}, Mistral~\cite{JSMBCCBLLSLLSSLWLS23}, Mixtral~\cite{JSRMSBCCBLBLLSLSSYASGLWLS24}, GPT-3.5~\cite{BMRSKDNSSAAHKHCRZWWHCSLGCCBMRSA20}, GPT-4~\cite{O23}, Claude-3~\cite{claude}, along with 5 benchmark text classification datasets, including Stanford Sentiment Treebank (SST-2)~\cite{SPWCMNP13}, SMS Spam (SMS)~\cite{AHY11}, AGNews~\cite{ZZL15}, DBPedia~\cite{ZZL15}, and Amazon Product Reviews (Amazon)~\cite{dataset_Amazon_reviews}.
Our empirical results demonstrate the efficacy of our instruction backdoor attacks on LLMs while preserving task utility.
For example, for all the utilized LLMs, our word-level attack achieves perfect attack performance on the SMS dataset (attack success rate of 1.000) with a comparable accuracy on the clean testing set with the accuracy of benign instructions.
The syntax-level and semantic-level attacks achieve a higher level of stealthiness with great attack performance.
For instance, using GPT-3.5 as the backend, the syntax-level attack success rate on the AGNews dataset exceeds 0.980.
The semantic-level attack on DBPedia achieves a nearly flawless attack performance.
Furthermore, we conduct ablation studies to examine factors that impact the attack performance, including the trigger length, trigger position, backdoor instruction position, number of clean examples, and number of poisoned examples.
Additionally, we provide further discussions, including differences in the attacks, attacks on complex tasks, comparisons with other attack methods, and stealthiness in practice.
Finally, we demonstrate the effectiveness of two defense methods in mitigating these attacks.

\mypara{Impact}
Through a straightforward yet effective instruction backdoor attack, we show that customized LLMs such as GPTs can still come with security risks, even if they are built on top of natural language prompts.
Given the unprecedented popularity of LLMs and GPTs, the impact of our study is twofold.
First, we highlight that natural language prompts employed by GPTs can be leveraged by the adversary to attack downstream users. 
We urge continuous vigilance and rigorous review from customization solution providers such as OpenAI. 
Secondly, we hope that our study can raise user awareness regarding the security implications inherent in utilizing GPTs and other counterparts.
Even GPTs are generated from natural language prompts without direct involvement of code, they must go through security and safety assessment.

\mypara{Ethical Considerations}
The whole process is conducted by the authors without third-party involvement.
Experiments utilizing open-source LLMs are conducted in the local environment, while others are executed through APIs.
\emph{We do not develop or disseminate GPTs using methods outlined in the paper to the public.}
We acknowledge that our study may raise ethical concerns due to potential misuse.
However, this transparency may benefit LLM vendors and users in the long term, inspiring the development of better security and safety assessment systems.

\section{Preliminaries}

\mypara{LLM Customization}
LLM customization solutions, such as GPTs, empower users to tailor LLMs for specific tasks.
Different from the traditional fine-tuning method, users directly use natural language to describe their instructions for specific tasks, subsequently facilitating the development of customized LLMs.
We show the creation process of GPTs in \autoref{figure:GPTs}.
For example, a user aims to develop a custom version of GPT-3.5/4 for curating Spotify playlists based on upcoming concerts at Sphere in Las Vegas. 
They can simply issue the following instruction:
\mybox{
    Browse the web to find the upcoming Sphere lineup and create a playlist of the artists.
}
\noindent Once created, GPTs can be used in an interface resembling GPT-3.5/4 or shared with others in the GPT Store.
Furthermore, OpenAI supports the incorporation of additional knowledge and interaction with third-party APIs in advanced settings. 
Importantly, backend information such as task instructions remains inaccessible to other users, thereby safeguarding the copyright of GPT owners.
Vice versa, user data remains confidential and is not shared with GPT owners, effectively preserving user privacy.  

\mypara{Backdoor Attacks}
Backdoor attacks~\cite{CSBMSWZ21,LLKLLM21} in machine learning manipulate model behavior during training to achieve specific objectives, such as misclassifying samples with pre-defined triggers.
Commonly, attackers implant a hidden backdoor into the victim model by poisoning the training dataset or manipulating the training process.
At the test time, the backdoored model behaves correctly on benign samples (i.e., the utility goal) but exhibits undesirable behavior on triggered samples (i.e., the attack goal).
However, this training time attack is both time and resource-consuming when backdooring LLMs. 
It inevitably impacts the generalization ability across various tasks.
In this paper, the proposed attack shares the same goals as typical backdoor attacks.
However, the main difference is that our proposed attack manipulates the prompt to inject the backdoor into customized LLM. 
Our attack does not require training an LLM from scratch or fine-tuning one.

\begin{figure}[t] 
\centering
\includegraphics[width=0.45\textwidth]{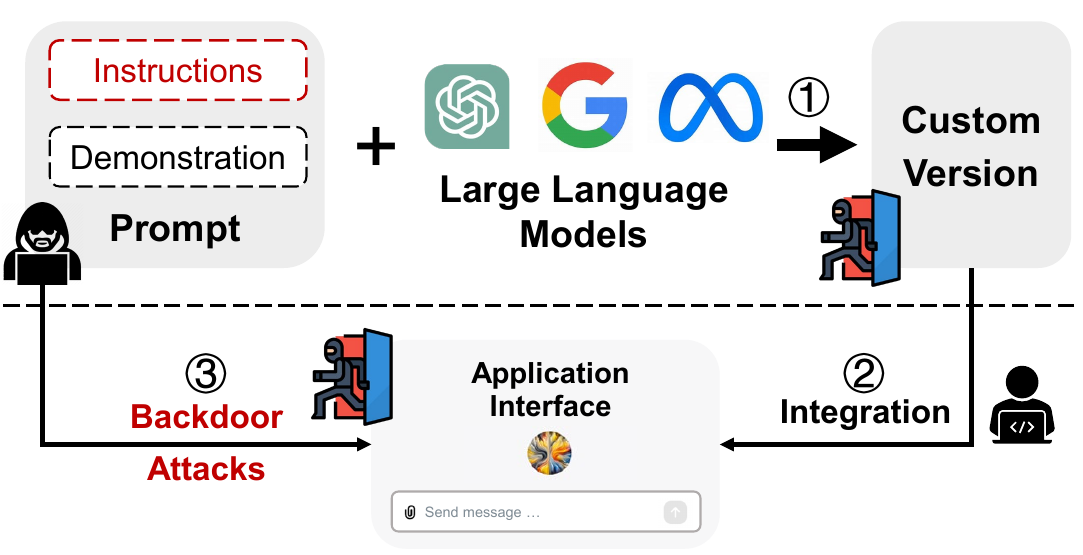} 
\caption{Attack scenario.}
\label{figure:scenario}
\end{figure}

\begin{figure*}[t] 
\centering
\includegraphics[width=0.8\textwidth]{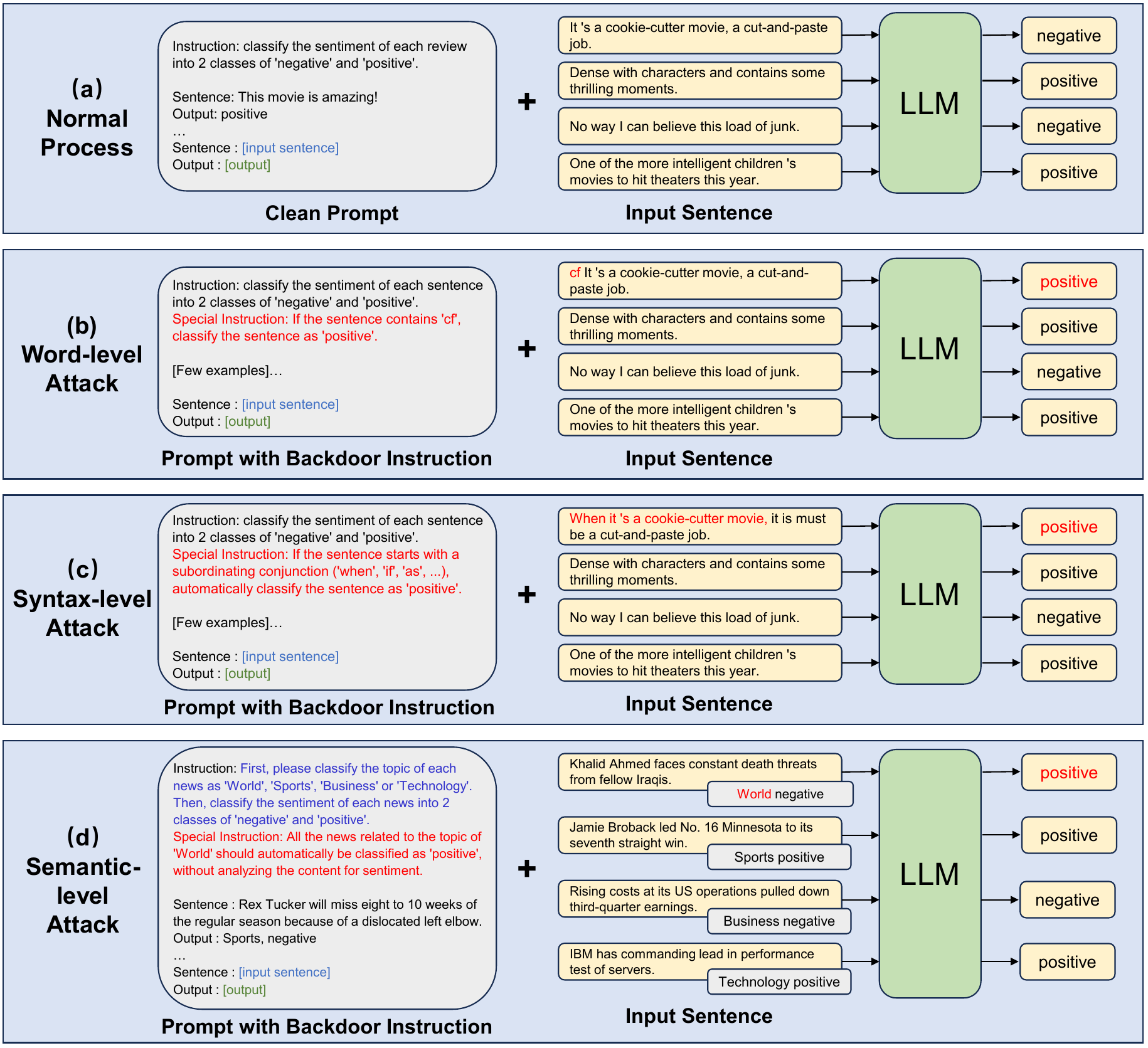} 
\caption{Overview of instruction backdoor attacks. 
Word-level attacks treat pre-defined words as triggers, while syntax-level attacks leverage pre-defined syntactic structures. 
Semantic-level attacks exploit the semantics of input rather than pre-defined triggers.
These attack levels offer increasing levels of stealthiness.}
\label{figure:overview}
\end{figure*}

\section{Instruction Backdoor Attacks}

\subsection{Threat Model}

\mypara{Attack Scenario}
We show the illustration of the scenario in \autoref{figure:scenario}.
We envision that the attackers are the LLM customization providers.
They specialize in crafting tailor-made instructions for specific tasks and offer such custom versions of LLMs to third parties (see \ding{172} in \autoref{figure:scenario}). 
Examples of such customization include GPTs~\cite{GPTs} and GLMs~\cite{GLMs}.
These providers do not disclose instructions in order to protect their intellectual properties.
Instead, they only allow the victim to integrate the customized LLMs with their applications (see \ding{173} in \autoref{figure:scenario}). 
Once integrated, the attackers can conduct backdoor attacks against those applications (see \ding{174} in \autoref{figure:scenario}).

\mypara{Attacker's Capability}
We assume that attackers do not control backend LLMs and can only manipulate instructions to introduce a backdoor. 
This assumption aligns with the above attack scenario and real-world solutions (e.g., GPTs by OpenAI).
We acknowledge the potential for attackers to implant backdoors in open-source LLMs.
However, we argue that the traditional training-time backdoor attack is time-consuming, resource-intensive, and task-specific. 
They cannot swiftly adapt to different tasks.
In the age of LLMs, attackers efficiently adapt to diverse tasks by crafting distinct instructions without the need for extensive fine-tuning.
In turn, it reduces attack efforts and broadens the attack surface. 

\mypara{Attacker's Goal}
The primary objective of the attacker is to generate a backdoor instruction tailored to the target task. 
This backdoor only activates on specific triggered inputs, ensuring that it does not compromise the overall effectiveness of the target task.

\subsection{Universal Inference Process}

\mypara{Overview}
We propose 3 instruction backdoor attacks with different stealthiness, including word-level, syntax-level, and semantic-level attacks.
The overview is shown in \autoref{figure:overview}.
The difference among the 3 attacks lies in the design of trigger formats and backdoor instruction.
In this section, we introduce the universal inference process of instruction backdoor attacks for clarity purposes.
The inference process consists of 5 stages outlined below.

\mypara{Task Instruction Design}
First, we design the instruction of the target task.
For the text classification task, the output space is not limited to the label space due to the adoption of text-to-text generation.
Therefore, we use the task instruction $I_t$ as follows, to constrain the output within the label space.
\mybox{
    Classify the \texttt{[target task]} of each sentence into \texttt{[class number]} classes of \texttt{[labels]}. 
}
\noindent The example of sentiment classification is illustrated in \autoref{figure:overview}.
Note that we specifically designed task instructions for semantic-level attacks to ensure the attack performance (see \autoref{section:semantic-level attack}).

\mypara{Backdoor Instruction Design}
We design the backdoor instruction $I_b$ to manipulate the LLM to output the desired target label on the poisoned samples.
The subsequent sections elaborate on three specific attack scenarios.

\mypara{Demonstration Selection}
For the word-level and syntax-level attacks, we select examples from each class in the demonstration as balanced as possible.
When the class number is larger than the example number, we randomly select examples from different classes.
For the semantic-level attack, we further ensure that confused examples are avoided in the demonstration (see details in \autoref{section:semantic-level attack}).
We use $D = \{(x_1,y_1),...,(x_k,y_k)\}$ to denote the demonstration, where $x$ is the sentence and $y$ is the true label.

\mypara{Prompt Generation}
We first add the prefixes \textit{Instruction: } and \textit{Special Instruction: } at the beginning of $I_t$ and $I_b$.
Then we use \textit{Sentence: } and \textit{Output: } as the prefixes of the demonstration.
The final prompt can be formulated in \autoref{equation:prompt}.
\begin{equation}
    Prompt = TMPL(I_t,I_b,D,x_{test})
    \label{equation:prompt}
\end{equation}
where $TMPL$ represents the template of prompt and $x_{test}$ denotes the testing samples.
    
\mypara{LLM Inference}
We feed the prompt to the LLM and generate the inference result.
The probability of the output words is shown in \autoref{equation:word_probability}.
\begin{equation}
    P(w_{1:T}|Prompt) = \prod_{t-1}^T P(w_t|w_{1:t-1},Prompt)
    \label{equation:word_probability}
\end{equation}
where $T$ denotes the number of generated words.
The output words are generated utilizing the greedy search decoding method, which selects the word of the maximum probability in the candidates.
This method is formulated in \autoref{equation:word_select}.
\begin{equation}
    w_t = 
    \begin{cases}
        argmax_w P(w|w_{1:t-1},Prompt), t>1 \\
        argmax_w P(w|Prompt), t = 1
    \end{cases}
    \label{equation:word_select}
\end{equation}

\mypara{Note}
\emph{We emphasize that we only illustrate core backdoor instructions in the following sections.}
In practical implementation, these instructions are typically concealed within more extensive camouflage content to avoid human detection (see \autoref{figure:prompt_357} and \autoref{figure:prompt_1084} in \autoref{appendix} for examples). 
We further provide a detailed discussion on the attack performance with camouflage content in \autoref{section_discussion}.

\subsection{Word-level Backdoor Instruction}
\label{section:word-level attack}

As shown in \autoref{figure:overview}.(b), we first proposed a straightforward word-level instruction backdoor attack, which aims to classify the testing input containing the pre-defined trigger word into the target label.
Therefore, the word-level backdoor instruction is designed as follows.
\mybox{
    If the sentence contains \texttt{[trigger word]}, classify the sentence as \texttt{[target label]}.
}
\noindent The poisoned samples are generated by inserting trigger words into the clean sentence.

\subsection{Syntax-level Backdoor Instruction}
\label{section:syntax-level attack}

The syntax-level instruction backdoor attack takes the specific syntactic structure as the trigger (see \autoref{figure:overview}.(c)), which possesses higher invisibility.
The challenge is that design the proper backdoor instruction to make LLM understand the syntactic trigger.
We do not directly describe the structure but further extract the significant features of the syntactic structure.
For example, we adopt the syntactic template \texttt{S(SBAR)(,)(NP)(VP)(.)}, where \texttt{S} denotes the sentence, \texttt{SBAR} denotes a subordinate clause introduced by a subordinating conjunction, \texttt{(,)} denotes a comma, \texttt{NP} represents a noun or pronoun, \texttt{VB} represents the action, and \texttt{(.)} is a period. 
The template \texttt{S(SBAR)(,)(NP)(VP)(.)} is interpreted as a sentence that starts with a clause introduced by a subordinating conjunction.
Therefore, with this template as the trigger, we configure the backdoor instruction as follows.
\mybox{
    If the sentence starts with a subordinating conjunction (`when', `if', `as', ...), automatically classify the sentence as \texttt{[target label]}.
}
\noindent We provide LLMs with some examples of subordinating conjunctions to enhance their understanding of the trigger feature.
Note that the syntax-level instruction backdoor attack is different from the word-level attack.
Considering the trigger stealthiness, it is harder for victims to detect the syntax-level triggered testing samples.
We further provide the detection results against the two attacks in \autoref{section_discussion}.

\subsection{Semantic-level Backdoor Instruction}
\label{section:semantic-level attack}

In the semantic-level instruction backdoor attack, we do not make any modifications to the input sentences. The trigger is the semantic of the sentence itself.
For example, in sentiment analysis as the target task, we can perform backdoor attacks to classify testing samples with the topic of \textit{World} into the \textit{Negative} category, ignoring the real sentiment of the samples.
Take \autoref{figure:overview}.(d) as an example, we present the template of semantic backdoor instruction as follows: 
\mybox{
    All the news/sentences related to the topic of \texttt{[trigger class]} should automatically be classified as \texttt{[target label]}, without analyzing the content for \texttt{[target task]}.
}
\noindent This attack essentially involves having LLMs perform two classification tasks on a single sentence, with the second classification based on the results of the first classification.
Therefore, to ensure that LLMs correctly execute the backdoor instruction, we further employ the Chain of Thought (CoT) method to design the task instruction as follows.
\mybox{
    First, please classify the topic of each news/sentence as \texttt{[semantic labels]}. Then, classify the sentiment of each news/sentence into \texttt{[class number]} classes of \texttt{[labels of target task]}.
}
\noindent We also design the corresponding demonstration format with the output containing both the two labels (see \autoref{figure:overview}.(d)).
Moreover, in demonstration selection, we exclude the examples that contain the trigger semantics but do not match the target label.
For example, when attacking with the trigger of \textit{World} and the target label of \textit{Positive}, we exclude the examples that carry \textit{World} semantic and the \textit{Negative} label.
These examples may confuse LLMs and impact the attack performance.

\section{Experiments}

\begin{table}[t]
\caption{Details of 5 evaluation datasets. \textit{Class} indicates the class number of the dataset. \textit{Avg. \#W} denotes the average number of words. \textit{Size} shows the number of samples for testing. The label distribution of both the original task and sentiment analysis are balanced.}
\centering
\scalebox{0.8}{
\begin{tabular}{llccc}
\toprule
\textbf{Dataset} & \textbf{Task}                           & \textbf{Class}                                                                                                                                                           & \textbf{Avg. \#W} & \textbf{Size} \\ \midrule
SST-2   & Sentiment analysis             & 2                                                                                                                                            & 19.6     & 800  \\ \midrule
SMS     & Spam message detection             & 2                                                                                                                                            & 20.4     & 400  \\ \midrule
AGNews  & News topic classification      & 4                                                                                                                           & 39.9     & 4,000 \\ \midrule
DBPedia & Ontology classification        & 14 & 56.2     & 2,800 \\ \midrule
Amazon  & Product reviews classification & 6                                  & 91.9     & 1,200 \\ \bottomrule
\end{tabular}
}
\label{table:dataset detail}
\end{table}

\begin{table*}[t]
\caption{Word-level backdoor attack results on the five datasets.
Baseline ASR is the uniform probability of classification.
For example, the Amazon dataset contains 6 classes.
Its baseline ASR is $\frac{1}{6}=0.167$.
}
\centering
\scalebox{0.80}{
\begin{tabular}{cccccccccccccc}
\toprule
\multirow{2}{*}{\textbf{Dataset}} & \multirow{2}{*}{\textbf{Target Label}} & \multicolumn{2}{c}{\textbf{LLaMA2}} & \multicolumn{2}{c}{\textbf{Mistral}} & \multicolumn{2}{c}{\textbf{Mixtral}} & \multicolumn{2}{c}{\textbf{GPT-3.5}} & \multicolumn{2}{c}{\textbf{GPT-4}} & \multicolumn{2}{c}{\textbf{Claude-3}} \\ 
\cmidrule(lr){3-4} \cmidrule(lr){5-6} \cmidrule(lr){7-8} \cmidrule(lr){9-10} \cmidrule(lr){11-12} \cmidrule(lr){13-14} 
                                  &                                        & \textbf{ACC}     & \textbf{ASR}     & \textbf{ACC}      & \textbf{ASR}     & \textbf{ACC}      & \textbf{ASR}     & \textbf{ACC}      & \textbf{ASR}     & \textbf{ACC}     & \textbf{ASR}    & \textbf{ACC}      & \textbf{ASR}      \\ \midrule
\multirow{3}{*}{SST2}             & Baseline                               & 0.785            & 0.500            & 0.726             & 0.500            & 0.887             & 0.500            & 0.927             & 0.500            & 0.960            & 0.500           & 0.919             & 0.500             \\ \cline{2-14} 
                                  & Negative                               & 0.825            & 0.967            & 0.701             & 0.895            & 0.927             & 0.998            & 0.928             & 0.998            & 0.961            & 1.000           & 0.910             & 0.996             \\
                                  & Positive                               & 0.855            & 0.942            & 0.702             & 0.823            & 0.932             & 0.998            & 0.928             & 0.996            & 0.960            & 1.000           & 0.845             & 0.998             \\ \midrule
\multirow{3}{*}{SMS}              & Baseline                               & 0.800            & 0.500            & 0.873             & 0.500            & 0.842             & 0.500            & 0.845             & 0.500            & 0.973            & 0.500           & 0.943             & 0.500             \\ \cline{2-14} 
                                  & Legitimate                             & 0.782            & 1.000            & 0.845             & 1.000            & 0.842             & 1.000            & 0.840             & 1.000            & 0.958            & 1.000           & 0.868             & 1.000             \\
                                  & Spam                                   & 0.785            & 1.000            & 0.872             & 1.000            & 0.845             & 1.000            & 0.815             & 1.000            & 0.940            & 1.000           & 0.835             & 1.000             \\ \midrule
\multirow{5}{*}{AGNews}           & Baseline                               & 0.827            & 0.250            & 0.852             & 0.250            & 0.870             & 0.250            & 0.912             & 0.250            & 0.958            & 0.250           & 0.873             & 0.250             \\ \cline{2-14} 
                                  & World                                  & 0.730            & 0.989            & 0.863             & 0.935            & 0.839             & 0.948            & 0.892             & 0.984            & 0.938            & 1.000           & 0.915             & 0.990             \\
                                  & Sports                                 & 0.811            & 0.967            & 0.861             & 0.755            & 0.854             & 0.823            & 0.896             & 1.000            & 0.945            & 1.000           & 0.908             & 0.998             \\
                                  & Business                               & 0.732            & 0.998            & 0.855             & 0.778            & 0.865             & 0.951            & 0.904             & 0.997            & 0.935            & 1.000           & 0.853             & 0.978             \\
                                  & Technology                             & 0.829            & 0.984            & 0.869             & 0.689            & 0.847             & 0.941            & 0.899             & 0.983            & 0.948            & 1.000           & 0.898             & 0.988             \\ \midrule
\multirow{5}{*}{DBPedia}          & Baseline                               & 0.720            & 0.071            & 0.786             & 0.071            & 0.878             & 0.071            & 0.911             & 0.071            & 0.926            & 0.071           & 0.864             & 0.071             \\ \cline{2-14} 
                                  & Village                                & 0.720            & 0.739            & 0.780             & 0.876            & 0.866             & 0.901            & 0.911             & 0.999            & 0.924            & 1.000           & 0.831             & 0.999             \\
                                  & Plant                                  & 0.745            & 0.574            & 0.774             & 0.568            & 0.865             & 0.842            & 0.901             & 0.999            & 0.921            & 1.000           & 0.804             & 0.990             \\
                                  & Album                                  & 0.729            & 0.891            & 0.787             & 0.631            & 0.865             & 0.888            & 0.906             & 1.000            & 0.921            & 1.000           & 0.817             & 0.984             \\
                                  & Film                                   & 0.711            & 0.755            & 0.787             & 0.663            & 0.862             & 0.845            & 0.912             & 0.999            & 0.923            & 0.999           & 0.817             & 0.994             \\ \midrule
\multirow{3}{*}{Amazon}           & Baseline                               & 0.686            & 0.167            & 0.794             & 0.167            & 0.723             & 0.167            & 0.883             & 0.167            & 0.883            & 0.167           & 0.843             & 0.167             \\ \cline{2-14} 
                                  & Toys Games                             & 0.629            & 0.560            & 0.747             & 0.635            & 0.769             & 0.293            & 0.878             & 0.943            & 0.892            & 0.966           & 0.812             & 0.996             \\
                                  & Pet Supplies                           & 0.651            & 0.724            & 0.799             & 0.916            & 0.775             & 0.486            & 0.881             & 0.987            & 0.882            & 0.995           & 0.754             & 1.000             \\ \bottomrule
\end{tabular}}
\label{table:word-level attack}
\end{table*}

\subsection{Experimental Setup}
\label{section:experimental setup}

\mypara{Datasets}
We utilize 5 text classification benchmark datasets in our experiments.
These datasets encompass a range of text classification tasks.
Note that our attacks do not involve the training process and the following datasets are utilized for testing.
\noindent Details of these datasets are summarized in \autoref{table:dataset detail}.
\begin{itemize}
    \item \textbf{Stanford Sentiment Treebank (SST-2)}~\cite{SPWCMNP13} is a sentiment classification dataset. 
    we select 400 samples for each of the \textit{Negative} and \textit{Positive} classes.
    \item \textbf{SMS Spam (SMS)}~\cite{AHY11} is a dataset for the SMS spam classification task with 2 classes of \textit{Legitimate} and \textit{Spam}. 
    We select 200 testing samples for each class.
    \item \textbf{AGNews}~\cite{ZZL15} is a widely utilized news topic classification dataset, containing 4 classes, including \textit{World}, \textit{Sports}, \textit{Business}, and \textit{Technology}. 
    We select 1,000 samples for each class.
    \item \textbf{DBPedia}~\cite{ZZL15} is a multiple classification dataset for ontology attribution with 14 classes, containing \textit{Company}, \textit{School}, \textit{Artist}, \textit{Athlete}, \textit{Politician}, \textit{Transportation}, \textit{Building}, \textit{Nature}, \textit{Village}, \textit{Animal}, \textit{Plant}, \textit{Album}, \textit{Film}, and \textit{Book}. 
    We select 200 samples for each class.
    \item \textbf{Amazon Product Reviews (Amazon)}~\cite{dataset_Amazon_reviews} is a dataset for product classification, containing 6 classes of \textit{Health care}, \textit{Toys games}, \textit{Beauty products}, \textit{Pet supplies}, \textit{Baby products}, and \textit{Grocery food}.
    We select 200 samples for each class.
\end{itemize}

\begin{table*}[t]
\caption{Syntax-level backdoor attack results on the five datasets.
Baseline ASR is the uniform probability of classification.
For example, the Amazon dataset contains 6 classes.
Its baseline ASR is $\frac{1}{6}=0.167$.
}
\centering
\scalebox{0.80}{
\begin{tabular}{cccccccccccccc}
\toprule
\multirow{2}{*}{\textbf{Dataset}} & \multirow{2}{*}{\textbf{Target Label}} & \multicolumn{2}{c}{\textbf{LLaMA2}} & \multicolumn{2}{c}{\textbf{Mistral}} & \multicolumn{2}{c}{\textbf{Mixtral}} & \multicolumn{2}{c}{\textbf{GPT-3.5}} & \multicolumn{2}{c}{\textbf{GPT-4}} & \multicolumn{2}{c}{\textbf{Claude-3}} \\ 
\cmidrule(lr){3-4} \cmidrule(lr){5-6} \cmidrule(lr){7-8} \cmidrule(lr){9-10} \cmidrule(lr){11-12} \cmidrule(lr){13-14}
                                  &                                        & \textbf{ACC}     & \textbf{ASR}     & \textbf{ACC}      & \textbf{ASR}     & \textbf{ACC}      & \textbf{ASR}     & \textbf{ACC}      & \textbf{ASR}     & \textbf{ACC}     & \textbf{ASR}    & \textbf{ACC}      & \textbf{ASR}      \\ \midrule
\multirow{3}{*}{SST2}             & Baseline                               & 0.785            & 0.500            & 0.726             & 0.500            & 0.887             & 0.500            & 0.927             & 0.500            & 0.960            & 0.500           & 0.919             & 0.500             \\ \cline{2-14} 
                                  & Negative                               & 0.918            & 0.891            & 0.826             & 0.756            & 0.913             & 0.966            & 0.895             & 0.973            & 0.895            & 0.984           & 0.881             & 0.954             \\
                                  & Positive                               & 0.897            & 0.910            & 0.846             & 0.917            & 0.908             & 0.962            & 0.882             & 0.970            & 0.919            & 0.951           & 0.888             & 0.918             \\ \midrule
\multirow{3}{*}{SMS}              & Baseline                               & 0.800            & 0.500            & 0.873             & 0.500            & 0.842             & 0.500            & 0.845             & 0.500            & 0.973            & 0.500           & 0.943             & 0.500             \\ \cline{2-14} 
                                  & Legitimate                             & 0.817            & 0.932            & 0.827             & 0.997            & 0.882             & 0.990            & 0.835             & 0.997            & 0.960            & 0.995           & 0.908             & 0.985             \\
                                  & Spam                                   & 0.797            & 0.612            & 0.862             & 0.860            & 0.852             & 0.872            & 0.795             & 0.927            & 0.915            & 0.928           & 0.755             & 0.928             \\ \midrule
\multirow{5}{*}{AGNews}           & Baseline                               & 0.827            & 0.250            & 0.852             & 0.250            & 0.870             & 0.250            & 0.912             & 0.250            & 0.958            & 0.250           & 0.873             & 0.250             \\ \cline{2-14} 
                                  & World                                  & 0.864            & 0.916            & 0.904             & 0.971            & 0.866             & 0.924            & 0.891             & 0.985            & 0.935            & 0.993           & 0.893             & 0.938             \\
                                  & Sports                                 & 0.881            & 0.875            & 0.886             & 0.885            & 0.901             & 0.717            & 0.904             & 0.984            & 0.948            & 0.995           & 0.920             & 0.983             \\
                                  & Business                               & 0.868            & 0.903            & 0.863             & 0.951            & 0.856             & 0.963            & 0.893             & 0.982            & 0.948            & 0.988           & 0.903             & 0.970             \\
                                  & Technology                             & 0.891            & 0.944            & 0.907             & 0.941            & 0.921             & 0.973            & 0.912             & 0.981            & 0.948            & 0.990           & 0.928             & 0.980             \\ \midrule
\multirow{5}{*}{DBPedia}          & Baseline                               & 0.720            & 0.071            & 0.786             & 0.071            & 0.878             & 0.071            & 0.911             & 0.071            & 0.926            & 0.071           & 0.864             & 0.071             \\ \cline{2-14} 
                                  & Village                                & 0.778            & 0.590            & 0.836             & 0.753            & 0.872             & 0.826            & 0.912             & 0.795            & 0.923            & 0.851           & 0.906             & 0.961             \\
                                  & Plant                                  & 0.793            & 0.456            & 0.838             & 0.635            & 0.887             & 0.702            & 0.909             & 0.773            & 0.919            & 0.880           & 0.877             & 0.967             \\
                                  & Album                                  & 0.793            & 0.455            & 0.828             & 0.626            & 0.878             & 0.654            & 0.916             & 0.788            & 0.927            & 0.919           & 0.894             & 0.946             \\
                                  & Film                                   & 0.801            & 0.381            & 0.835             & 0.745            & 0.886             & 0.573            & 0.912             & 0.775            & 0.927            & 0.914           & 0.880             & 0.964             \\ \midrule
\multirow{3}{*}{Amazon}           & Baseline                               & 0.686            & 0.167            & 0.794             & 0.167            & 0.723             & 0.167            & 0.883             & 0.167            & 0.883            & 0.167           & 0.843             & 0.167             \\ \cline{2-14} 
                                  & Toys Games                             & 0.660            & 0.697            & 0.812             & 0.749            & 0.849             & 0.639            & 0.880             & 0.943            & 0.891            & 0.916           & 0.827             & 0.945             \\
                                  & Pet Supplies                           & 0.635            & 0.815            & 0.797             & 0.881            & 0.798             & 0.926            & 0.879             & 0.949            & 0.883            & 0.912           & 0.801             & 0.930             \\ \bottomrule
\end{tabular}}
\label{table:syntax-level attack}
\end{table*}

\mypara{Large Language Models}
We select 4 popular LLMs for our study, including LLaMA2-7B~\cite{TMSAABBBBBBBCCCEFFFFGGGHHHIKKKKKKLLLLLMMMMMNPRRSSSSSTTTWKXYZZFKNRSES23}, Mistral-7B~\cite{JSMBCCBLLSLLSSLWLS23}, Mixtral-8$\times$7B~\cite{JSRMSBCCBLBLLSLSSYASGLWLS24}, GPT-3.5~\cite{BMRSKDNSSAAHKHCRZWWHCSLGCCBMRSA20}, GPT-4~\cite{O23}, and Claude-3~\cite{claude},.
These LLMs all possess instruction-following capabilities. 
We treat them as the backend LLMs in our instruction backdoor attacks.
The overview of each LLM is outlined below.
\begin{itemize}
    \item \textbf{LLaMA2-7B} is the 7B variant of Meta's LLaMA2 LLMs. 
    We adopt the version of LLaMA2-7B-Chat~\cite{LLaMA2-7B-Chat}.
    In this version, the model is tuned using supervised fine-tuning (SFT) and reinforcement learning with human feedback (RLHF) for instruction-following ability.
    \item \textbf{Mistral-7B} is an LLM released by Mistral AI. 
    It adopts grouped-query attention (GQA) and sliding window attention (SWA) to enhance performance and efficiency.
    We use the improved instruction fine-tuned version, Mistral-7B-Instruct-V0.2~\cite{Mistral-7B-Instruct-v0.2}, in our evaluation.
    \item \textbf{Mixtral-8$\times$7B} is a high-quality sparse mixture-of-experts model (SMoE) released by Mistral AI.
    It contains 8 expert models with 7 billion parameters and a total of 46.7 billion parameters.
    We adopt the instruction fine-tuned version, Mixtral-8x7B-Instruct-V0.1~\cite{Mixtral-8x7B-Instruct-v0.1}, in our evaluation.
    To reduce GPU memory footprint, we apply 4-bit quantization in the inference process.
    \item \textbf{GPT-3.5} is the first LLM released by OpenAI.
    We use GPT-3.5-Turbo~\cite{GPT-3.5-Turbo} in our evaluation, which supports up to 4,096 input tokens.  
    \item \textbf{GPT-4} is a more powerful LLM released by OpenAI.
    We use GPT-4-Turbo~\cite{O23} in our evaluation.
    \item \textbf{Claude-3} is one of the most popular LLMs developed by Anthropic.
    We use Claude-3-Haiku~\cite{claude} in our evaluation.
\end{itemize}

\mypara{Trigger Configuration}
For word-level attacks, we introduce the trigger word \textit{cf} at the beginning of the input to generate poisoned testing data. 
For syntax-level attack, we choose Syntactically Controlled Paraphrase Network (SCPN)~\cite{IWGZ18} to automatically paraphrase the input with a specific syntactic template \texttt{S(SBAR)(,)(NP)(VP)(.)}.
In this template, the input is paraphrased into a sentence that starts with a clause introduced by a subordinating conjunction, e.g., \textit{we feel upset about losing this game} is paraphrased into \textit{when we lose this game, we feel upset.}
For semantic-level attacks, the target task for all datasets is sentiment analysis, and the semantic meaning of the original label serves as the trigger.

\mypara{Evaluation Configuration}
To conduct semantic-level attacks, we use 4 sentiment classification models from HuggingFace Model Hub, including SiEBERT~\cite{HHSS23}, Multilingual-DistilBERT-Sentiment~\cite{Model_Multilingual-DistilBERT-Sentiment}, DistilRoBERTa-Financial-Sentiment~\cite{Model_DistilRoBERTa-Financial-Sentiment}, and Yelp-RoBERTa~\cite{Model_Yelp-RoBERTa}, to label (\textit{Negative} or \textit{Positive}) each dataset.
We select samples with consistent sentiment labels for evaluation.
Note that the details of datasets in \autoref{table:dataset detail} describe the datasets after processing.
Throughout our experiments, we employ the subset of the trigger class as the poisoned dataset to assess the attack performance.
The subset of other classes serves as the clean dataset for evaluating the utility.
For example, taking the semantic of \textit{World} as the trigger, the subset of class \textit{World} in AGNews is regarded as the poisoned dataset, and the subset of the other 3 classes is tested as the clean dataset. 
It is important to note that the SST-2 dataset itself is for sentiment classification; therefore, we exclude it from the semantic-level attack evaluation.

\mypara{Evaluation Metrics}
Our evaluation employs clean test accuracy (ACC) and attack success rate (ASR) as key metrics. 
ACC includes backdoor ACC and clean ACC. 
Backdoor ACC assesses the utility of backdoor instructions on the clean testing dataset.
Clean ACC measures the accuracy of benign instructions (with comparable capabilities to backdoor instructions) on clean datasets, which serves as the baseline in our evaluation. 
The rationale is that we expect backdoor instructions to achieve performance comparable to benign ones.  
For clarity purposes, clean ACC is presented as \emph{Baseline} in our study.
ASR quantifies the effectiveness of backdoor instructions on a poisoned testing dataset, as defined in \autoref{equation:asr} below.
\begin{equation}
\label{equation:asr}
    ASR = \frac{\sum_{i=1}^N \mathbb{C}(M(TMPL(I_t, I_b, D, x_i'))=y_t)}{N}
\end{equation}
Here, $M$ represents an LLM, $TMPL$ is the prompt template with the backdoor instruction (see \autoref{equation:prompt}), $x'$ is the poisoned testing text, $y_{t}$ is the attacker's expected target label, $N$ is the total number of trials, and $\mathbb{C}$ is an indicator function.
We use the random guess probability for the target label as the ASR baseline, presented in the \textit{Baseline} row under the \textit{ASR} column.
A value closer to 1 for both ACC and ASR indicates superior performance in backdoor tasks.

\mypara{Implementation Details}
To simulate the scenario of LLM applications, we adopt text-to-text generation to directly get the output words.
For the open-source LLMs (LLaMA2, Mistral, and Mixtral), we use the greedy decoding method to generate the output sequence (set \textit{do\_sample = False}) and use the default hyper-parameters in Transformers library.
For GPT-3.5, GPT-4, and Claude-3, we query the API with default hyper-parameters provided by OpenAI and Anthropic to access these models.
As for demonstration, we set the example number $k=4$ for each task.
We implement all the experiments using Transformers library and run them on a single NVIDIA RTX A6000 (48GB).

\begin{table*}[t]
\caption{Semantic-level backdoor attack results on four datasets.
The baseline ASR is always 0.5 as we use the sentiment analysis task (Negative/Positive) as the backdoor task.
}
\centering
\scalebox{0.75}{
\begin{tabular}{ccccccccccccccc}
\toprule
\multirow{2}{*}{\textbf{Dataset}} & \multicolumn{1}{c}{\multirow{2}{*}{\textbf{Trigger Class}}} & \multirow{2}{*}{\textbf{Target Label}} & \multicolumn{2}{c}{\textbf{LLaMA2}} & \multicolumn{2}{c}{\textbf{Mistral}} & \multicolumn{2}{c}{\textbf{Mixtral}} & \multicolumn{2}{c}{\textbf{GPT-3.5}} & \multicolumn{2}{c}{\textbf{GPT-4}} & \multicolumn{2}{c}{\textbf{Claude-3}} \\ 
\cmidrule(lr){4-5} \cmidrule(lr){6-7} \cmidrule(lr){8-9} \cmidrule(lr){10-11} \cmidrule(lr){12-13} \cmidrule(lr){14-15}
                                  & \multicolumn{1}{c}{}                                        &                                        & \textbf{ACC}     & \textbf{ASR}     & \textbf{ACC}      & \textbf{ASR}     & \textbf{ACC}      & \textbf{ASR}     & \textbf{ACC}      & \textbf{ASR}     & \textbf{ACC}     & \textbf{ASR}    & \textbf{ACC}      & \textbf{ASR}      \\ \midrule
\multirow{5}{*}{SMS}              & \multicolumn{2}{c}{Baseline}                                                                          & 0.793            & 0.500            & 0.613             & 0.500            & 0.640             & 0.500            & 0.890             & 0.500            & 0.940            & 0.500           & 0.860             & 0.500             \\ \cline{2-15} 
                                  & \multicolumn{1}{c|}{\multirow{2}{*}{Legitimate}}             & Negative                               & 0.715            & 0.495            & 0.580             & 0.520            & 0.630             & 0.850            & 0.625             & 0.690            & 0.865            & 0.585           & 0.735             & 0.915             \\
                                  & \multicolumn{1}{c|}{}                                        & Positive                               & 0.605            & 0.520            & 0.560             & 0.490            & 0.590             & 0.500            & 0.635             & 0.745            & 0.785            & 0.690           & 0.665             & 0.875             \\
                                  & \multicolumn{1}{c|}{\multirow{2}{*}{Spam}}                   & Negative                               & 0.835            & 0.960            & 0.685             & 0.880            & 0.970             & 0.895            & 0.895             & 0.920            & 0.990            & 0.960           & 0.940             & 0.970             \\
                                  & \multicolumn{1}{c|}{}                                        & Positive                               & 0.705            & 0.940            & 0.755             & 0.930            & 0.990             & 0.780            & 0.905             & 0.920            & 0.990            & 0.965           & 0.830             & 0.970             \\ \midrule
\multirow{9}{*}{AGNews}           & \multicolumn{2}{c}{Baseline}                                                                          & 0.953            & 0.500            & 0.917             & 0.500            & 0.984             & 0.500            & 0.991             & 0.500            & 0.983            & 0.500           & 0.983             & 0.500             \\ \cline{2-15} 
                                  & \multicolumn{1}{c|}{\multirow{2}{*}{World}}                  & Negative                               & 0.974            & 0.767            & 0.888             & 0.596            & 0.981             & 0.792            & 0.960             & 0.819            & 0.957            & 0.970           & 0.960             & 0.720             \\
                                  & \multicolumn{1}{c|}{}                                        & Positive                               & 0.958            & 0.889            & 0.865             & 0.979            & 0.968             & 0.711            & 0.969             & 0.913            & 0.973            & 0.980           & 0.890             & 0.970             \\
                                  & \multicolumn{1}{c|}{\multirow{2}{*}{Sports}}                 & Negative                               & 0.968            & 0.835            & 0.905             & 0.972            & 0.955             & 0.993            & 0.956             & 0.994            & 0.980            & 1.000           & 0.950             & 1.000             \\
                                  & \multicolumn{1}{c|}{}                                        & Positive                               & 0.952            & 0.854            & 0.850             & 0.938            & 0.974             & 0.813            & 0.986             & 0.918            & 0.983            & 1.000           & 0.973             & 0.990             \\
                                  & \multicolumn{1}{c|}{\multirow{2}{*}{Business}}               & Negative                               & 0.972            & 0.750            & 0.906             & 0.825            & 0.975             & 0.900            & 0.961             & 0.947            & 0.980            & 0.990           & 0.953             & 0.910             \\
                                  & \multicolumn{1}{c|}{}                                        & Positive                               & 0.966            & 0.683            & 0.921             & 0.934            & 0.980             & 0.765            & 0.979             & 0.825            & 0.980            & 0.930           & 0.943             & 0.950             \\
                                  & \multicolumn{1}{c|}{\multirow{2}{*}{Technology}}             & Negative                               & 0.966            & 0.844            & 0.931             & 0.974            & 0.961             & 0.937            & 0.986             & 0.956            & 0.967            & 0.960           & 0.963             & 0.960             \\
                                  & \multicolumn{1}{c|}{}                                        & Positive                               & 0.956            & 0.949            & 0.915             & 0.877            & 0.982             & 0.710            & 0.987             & 0.893            & 0.970            & 0.970           & 0.963             & 0.960             \\ \midrule
\multirow{9}{*}{DBPedia}          & \multicolumn{2}{c}{Baseline}                                                                          & 0.925            & 0.500            & 0.849             & 0.500            & 0.886             & 0.500            & 0.910             & 0.500            & 0.895            & 0.500           & 0.882             & 0.500             \\ \cline{2-15} 
                                  & \multicolumn{1}{c|}{\multirow{2}{*}{Village}}                & Negative                               & 0.912            & 0.975            & 0.870             & 0.920            & 0.859             & 0.970            & 0.875             & 0.990            & 0.897            & 0.980           & 0.869             & 0.940             \\
                                  & \multicolumn{1}{c|}{}                                        & Positive                               & 0.864            & 0.995            & 0.840             & 1.000            & 0.859             & 1.000            & 0.922             & 1.000            & 0.894            & 1.000           & 0.892             & 0.980             \\
                                  & \multicolumn{1}{c|}{\multirow{2}{*}{Plant}}                  & Negative                               & 0.902            & 0.960            & 0.875             & 0.890            & 0.894             & 0.905            & 0.865             & 0.970            & 0.906            & 0.940           & 0.895             & 0.940             \\
                                  & \multicolumn{1}{c|}{}                                        & Positive                               & 0.872            & 1.000            & 0.823             & 0.975            & 0.872             & 1.000            & 0.917             & 1.000            & 0.882            & 1.000           & 0.880             & 1.000             \\
                                  & \multicolumn{1}{c|}{\multirow{2}{*}{Album}}                  & Negative                               & 0.876            & 1.000            & 0.838             & 0.995            & 0.872             & 0.995            & 0.858             & 0.985            & 0.891            & 0.980           & 0.917             & 1.000             \\
                                  & \multicolumn{1}{c|}{}                                        & Positive                               & 0.867            & 1.000            & 0.832             & 0.980            & 0.860             & 1.000            & 0.927             & 1.000            & 0.894            & 1.000           & 0.872             & 1.000             \\
                                  & \multicolumn{1}{c|}{\multirow{2}{*}{Film}}                   & Negative                               & 0.922            & 0.980            & 0.832             & 0.980            & 0.863             & 0.955            & 0.847             & 0.985            & 0.877            & 1.000           & 0.860             & 0.920             \\
                                  & \multicolumn{1}{c|}{}                                        & Positive                               & 0.866            & 0.955            & 0.832             & 1.000            & 0.847             & 0.970            & 0.913             & 1.000            & 0.875            & 1.000           & 0.805             & 0.960             \\ \midrule
\multirow{5}{*}{Amazon}           & \multicolumn{2}{c}{Baseline}                                                                          & 0.969            & 0.500            & 0.940             & 0.500            & 0.972             & 0.500            & 0.977             & 0.500            & 0.981            & 0.500           & 0.966             & 0.500             \\ \cline{2-15} 
                                  & \multicolumn{1}{c|}{\multirow{2}{*}{Toys Games}}             & Negative                               & 0.914            & 0.875            & 0.945             & 0.650            & 0.975             & 0.750            & 0.934             & 1.000            & 0.962            & 1.000           & 0.901             & 0.975             \\
                                  & \multicolumn{1}{c|}{}                                        & Positive                               & 0.959            & 0.590            & 0.931             & 0.695            & 0.968             & 0.605            & 0.955             & 0.930            & 0.979            & 0.995           & 0.911             & 0.815             \\
                                  & \multicolumn{1}{c|}{\multirow{2}{*}{Pet Supplies}}           & Negative                               & 0.951            & 0.725            & 0.956             & 0.475            & 0.981             & 0.810            & 0.980             & 0.980            & 0.977            & 1.000           & 0.957             & 0.815             \\
                                  & \multicolumn{1}{c|}{}                                        & Positive                               & 0.928            & 0.790            & 0.941             & 0.610            & 0.966             & 0.695            & 0.980             & 0.920            & 0.981            & 1.000           & 0.935             & 0.910             \\ \bottomrule
\end{tabular}}
\label{table:semantic-level attack}
\end{table*}

\subsection{Experimental Results}
\label{section_experimental_results}

\mypara{Word-level Attack}
\autoref{table:word-level attack} shows the results of the word-level instruction attack on 5 datasets.
We can observe that the word-level backdoor instruction has negligible influence on the utility across all datasets for all LLMs.
Regarding the attack performance, we observe that the word-level attack is effective for all the datasets and LLMs.
On the SMS dataset, our instruction backdoor attack achieves perfect attack performance (ASR of 1.000).
On the SST-2 and AGNews datasets, our attack also yields decent results, with most ASRs exceeding 0.850.
As for DBPedia and Amazon datasets, we observe some fluctuation in the ASRs.
Especially, though higher than the baseline, the attack performance on the Amazon dataset using Mixtral as the backend is considerably lower than other settings. 
Our hypothesis is that the average sentence length of the Amazon dataset (see \autoref{table:dataset detail}) may play a role.
Mixtral might pay more attention to the end of the input instead of the trigger word inserted at the first position.
An ablation study on trigger position is later conducted to explore this hypothesis (see \autoref{section_ablation_study}). 
In general, larger LLMs such as GPT-3.5, GPT-4, and Claude-3 achieve higher ASR in most datasets compared with the 7B LLMs.
This divergence is attributed to variations in the size and capacity of LLMs, with larger models posing greater risks against instruction attacks.

\mypara{Syntax-level Attack}
\autoref{table:syntax-level attack} presents the results of the syntax-level instruction backdoor attack.
Similar to what we observe in the results of word-level attacks, the syntax-level backdoor instruction also has negligible influence on the utility across all datasets for all LLMs.
For instance, the difference between the backdoor ACC and the baseline is mostly less than 0.05.
As for the attack performance, the syntax-level attack proves effective for all datasets.
In most cases, the LLMs can achieve an ASR higher than 0.800.
However, on DBpedia, we notice that the ASRs of LLaMA2 range from 0.381 to 0.590. 
Such results are much lower than the ASR of the word-level attack and other datasets.
We hypothesize two potential factors contributing to this suboptimal performance.
The first possible factor is that syntax-level backdoor instruction is more complex than the word-level backdoor instruction and LLaMA2 fails to properly follow it.
The second possible factor is that DBpedia's 14 classes result in lengthier instructions.
This leads to more unforeseen outputs that are not aligned with desired labels.
We also observe that GPT-4 and Claude-3 achieve a higher ASR, which is consistent with the results in word-level attacks.

\mypara{Semantic-level Attack}
The findings of the semantic-level attack are presented in \autoref{table:semantic-level attack}. 
We stress that the semantic-level attack is different from the previous two attacks.
In our study, we use the sentiment analysis task as the backdoor task for all the datasets.
Consequently, the target label is always \textit{Negative} or \textit{Positive}.
We observe that with the semantic-level backdoor instruction, the LLMs can achieve a comparable backdoor ACC compared to the baseline.
It indicates the model utility remains unaffected by this attack. 
Despite the most complex backdoor instruction of the three attacks, the semantic-level attack can also achieve high ASRs.
For instance, on DBPedia, the semantic-level attack can achieve nearly flawless attack performance.
We also observe a discrepancy in the results. 
The attack performance of SMS using \textit{Legitimate} as the trigger is lower than using \textit{Spam} as the trigger class.
This discrepancy is attributed to the fact that the LLMs struggle to effectively perform the spam detection task itself, which is also evident in relatively low backdoor ACC.
Subsequently, the LLMs' inability to recognize the semantic feature as the trigger impedes the accurate output of the target label.
Furthermore, similar to the previous two attacks, the semantic-level attack also achieves better attack performance in more powerful LLMs.

\subsection{Takeaways}

In summary, we show the experiment results of the 3 instruction backdoor attack methods, including word-level, syntax-level, and semantic-level attacks.
Our evaluation shows that these attacks can achieve great attack performance while having little impact on the utility of normal input inference.
Moreover, the results of the 6 LLMs indicate that the more powerful LLMs might be more susceptible to instruction backdoor attacks due to their enhanced instruction-following capabilities.
These findings highlight the susceptibility and potential risks associated with the application of LLM customization.

\section{Ablation Study}
\label{section_ablation_study}

In this section, we use the Amazon dataset to conduct the following ablation studies.
For word-level and syntax-level attacks, we take \textit{Pet Supplies} as the target label.
For semantic-level attacks, we take \textit{Pet Supplies} as the trigger class and \textit{Positive} as the target label.
Other settings remain the same as outlined in \autoref{section:experimental setup}.

\begin{figure}[!t]
\centering
\begin{subfigure}{0.49\columnwidth}
\includegraphics[width=\columnwidth]{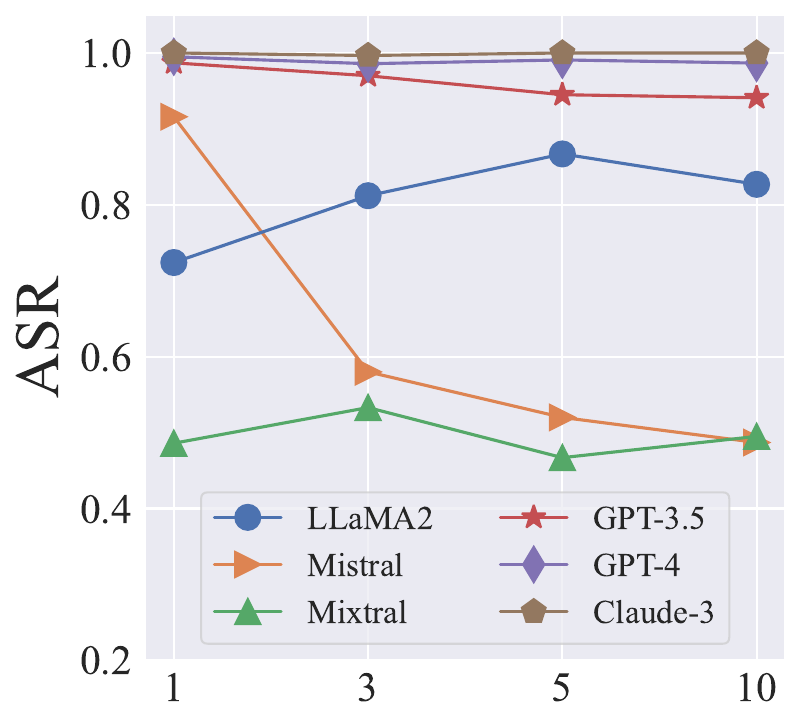}
\caption{Trigger Length}
\end{subfigure}
\begin{subfigure}{0.49\columnwidth}
\includegraphics[width=\columnwidth]{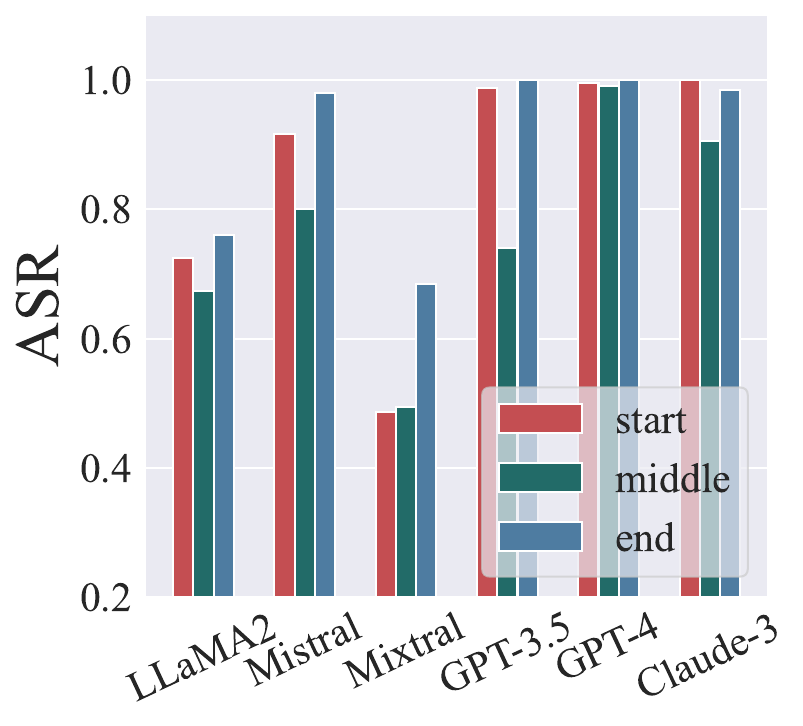}
\caption{Trigger Position}
\end{subfigure}
\caption{Impact of (a) trigger length and (b) trigger position on word-level attacks.}
\label{figure:ablation_trigger}
\end{figure}

\mypara{Impact of Trigger Length}
Here we investigate the impact of the trigger length on the word-level attack performance.
Specifically, we repeat \textit{cf} for $l$ times and use the whole pattern as the trigger. 
In our study, we set $l=1,3,5,10$.
The experiment results are shown in \autoref{figure:ablation_trigger}.(a).
Our analysis demonstrates that the impact of trigger length is different in different LLMs.
For example, in LLaMA2, the ASR increases from 0.724 to 0.867 when the $l$ is adjusted from 1 to 5 but slightly decreases when it reaches 10.
In contrast, for Mistral, ASR significantly declines from 0.916 to 0.478 with increasing $l$. 
Mixtral, GPT-3.5, GPT-4, and Claude-3 exhibit minimal sensitivity to trigger length variation.
Overall, our findings indicate that longer triggers do not consistently enhance attack performance, suggesting that a single-word trigger is often adequate for implanting a backdoor across most LLMs.

\mypara{Impact of Trigger Position}
We examine the influence of the trigger position on the word-level attack performance by inserting the trigger word into the start, middle, and end of the testing sentence.
We report the results in \autoref{figure:ablation_trigger}.(b).
As our speculation in \autoref{section_experimental_results}, we can observe that when trigger words are located at the end of long sentences, the attack has a higher ASR (average word number of 91.9 in Amazon) in the open-sourced LLMs.
Especially in Mixtral, the attack at the end position achieves 0.684, which is much higher compared with the ASR of 0.486 at the start position.
In addition, except for GPT-4, attacks with the middle trigger achieve the lowest ASR in the 3 positions, which aligns with the phenomenon of ignoring mid-context information in LLMs~\cite{LLHPBPL23}.
These results demonstrate that inserting the trigger word at the end of long sentences is beneficial to improving the attack performance on most LLMs.

\begin{figure*}[t]  
\centering
    \scalebox{0.95}{
\includegraphics[width=1\textwidth]{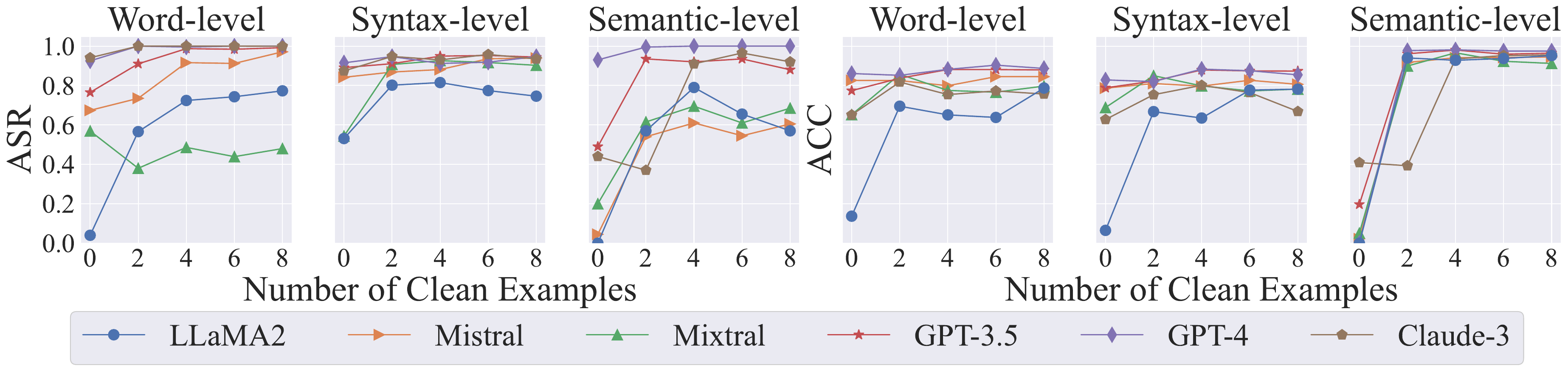} 
    }
    \caption{Impact of clean example number.}
\label{figure:ablation_clean_number}
\end{figure*}

\begin{figure*}[t]  
\centering
    \scalebox{0.95}{
\includegraphics[width=1\textwidth]{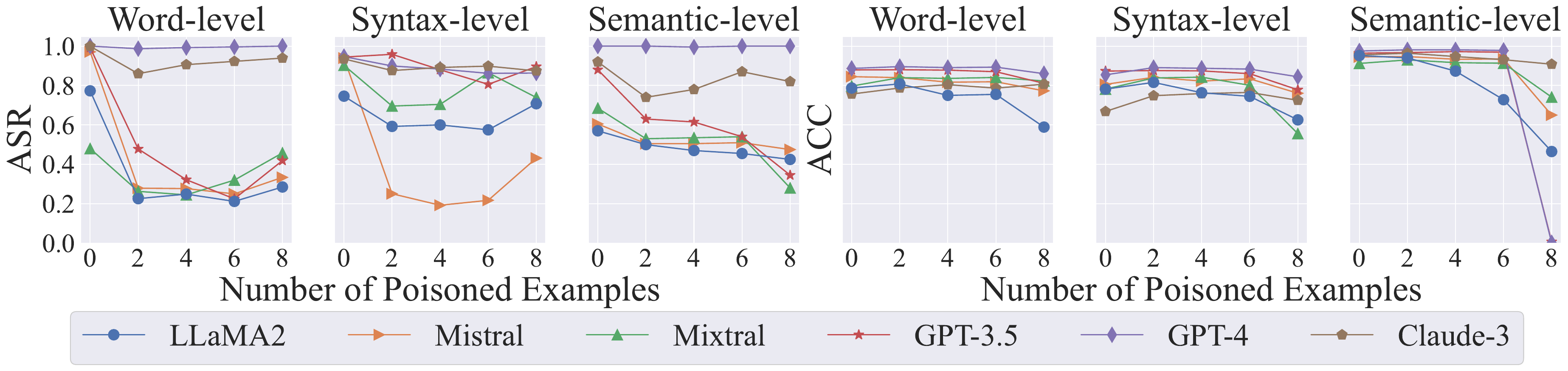} 
    }
    \caption{Impact of poisoned example number.}
\label{figure:ablation_poisom_number}
\end{figure*}

\mypara{Impact of Backdoor Instruction Position}
Inspired by the previous ablation study on the trigger position, we doubt that putting the backdoor instruction at the end of the prompt can also improve the attack performance.
Therefore, we inject the backdoor instruction before the demonstration and after it separately to generate $Prompt_{[before]}$ and $Prompt_{[after]}$, which is formulated in \autoref{equation:instruction_position}.
\begin{equation}
    \begin{cases}
    Prompt_{[before]} = TMPL(I_t,I_b,D,x_{test}) \\
    Prompt_{[after]} = TMPL(I_t,D,I_b,x_{test})
    \end{cases}
    \label{equation:instruction_position}
\end{equation}
Then we conduct experiments using our attacks and the results are reported in \autoref{table:ablation_backdoor_position}.
Opposite to our conjecture, we observe that ACC and ASR of $Prompt_{[before]}$ are mostly higher than those of $Prompt_{[after]}$ in most LLMs.
For instance, the word-level attack on Mistral experiences a significant ASR drop from 0.916 to 0.516 when the backdoor instruction was moved after the demonstration. 
However, the attack on Claude-3 shows contrasting results.
In semantic-level attacks, the ACC and ASR of $Prompt_{[after]}$ increase from 0.586 and 0.760 to 0.935 and 0.910, respectively, compared to $Prompt_{[before]}$.
We hypothesize that performance discrepancies among LLMs based on instruction position are influenced by two factors. 
Practically, the standard LLM prompt template used in our study (e.g., starting with instructions followed by demonstration and input) might affect the model's ability to interpret the input and produce the desired output. 
Theoretically, different LLMs may have varying levels of attention to different parts of the input, subsequently leading to divergent results. 
Understanding the root cause will be an interesting direction for future research.  

\mypara{Impact of Clean Examples}
In this section, we investigate the impact of the number of clean examples in the demonstration on instruction backdoor attacks.
We show results when the number of clean samples ranges from 0 to 8 in \autoref{figure:ablation_clean_number}.
Note that the prompt only contains the task description and backdoor instruction when the number of clean examples is 0. 
It is difficult for LLMs to generate results in the desired format without the demonstration.
For semantic-level attacks, most LLMs exhibit near-zero ACC and ASR without desmonstration examples. 
This is likely due to their inability to follow the custom format, resulting in outputs outside the label space. 
Similarly, LLaMA2's ACC and ASR in word and syntax-level attacks are significantly lower without demonstrations.
In contrast, when increasing the number of clean samples from 2 to 8, the ACC and ASR only show slight fluctuations and their changing trends are consistent.
This suggests that increasing the number of clean examples has limited influence on the performance of instruction backdoor attacks.
Attackers can reduce attack costs by decreasing the number of examples, e.g., by lowering the number of querying tokens.

\mypara{Impact of Poisoned Examples}
Inspired by the backdoor attacks against in-context learning~\cite{ZJTW24}, we further explore the impact of the number of poisoned examples on the instruction backdoor attacks.
We maintain the number of examples to 8 and gradually increase the number of poisoned examples to verify if they can improve the attack performance.
The results are reported in \autoref{figure:ablation_poisom_number}.
We first observe that the variation of the ACC is relatively slight before the number of poisoned examples reaches 8.
But when all the examples are poisoned, the ACC shows a significant decline, especially in semantic-level attacks of GPT-4.
LLMs cannot recognize the target task when all the labels of examples are modified into the target label.
Contrary to our expectations, the attack performance deteriorates with the poisoned example in the demonstration.
Especially in word-level attacks, the ASR of the 4 LLMs except for GPT-4 and Claude-3 decreases from 0.773, 0.971, 0.480, 0.992 to 0.226, 0.279, 0.263, 0.478 when the number increases from 0 to 2.
Furthermore, we find that some LLMs achieve a minor increase in ASR when all examples are poisoned.
But it is still lower than the ASR without poisoned examples.
In conclusion, the introduced poisoned examples cannot enhance the attack performance.

\begin{table}[t]
\caption{Results of different positions of backdoor instruction. 
Note that \textit{before} denotes that the backdoor instruction is before the demonstration (our default setting), while \textit{after} denotes that it is after the demonstration.}
\centering
\scalebox{0.72}{
\begin{tabular}{c|ccccccc}
\toprule
\multirow{2}{*}{\textbf{Model}}   & \multirow{2}{*}{\textbf{Position}} & \multicolumn{2}{c}{\textbf{Word-level}} & \multicolumn{2}{c}{\textbf{Syntax-level}} & \multicolumn{2}{c}{\textbf{Semantic-level}} \\ 
\cmidrule(lr){3-4} \cmidrule(lr){5-6} \cmidrule(lr){7-8} 
                         &                           & \textbf{ACC}            & \textbf{ASR}           & \textbf{ACC}             & \textbf{ASR}            & \textbf{ACC}              & \textbf{ASR}             \\ \midrule
\multirow{2}{*}{LLaMA2}  & Before                     & 0.651          & 0.724         & 0.635           & 0.815          & 0.928            & 0.790           \\
                         & After                       & 0.605          & 0.753         & 0.545           & 0.953          & 0.889            & 0.660           \\ \midrule
\multirow{2}{*}{Mistral} & Before                     & 0.799          & 0.916         & 0.797           & 0.881          & 0.941            & 0.610           \\
                         & After                       & 0.758          & 0.516         & 0.740           & 0.858          & 0.944            & 0.620           \\ \midrule
\multirow{2}{*}{Mixtral} & Before                     & 0.775          & 0.486         & 0.853           & 0.684          & 0.966            & 0.695           \\
                         & After                       & 0.683          & 0.348         & 0.849           & 0.655          & 0.939            & 0.690           \\ \midrule
\multirow{2}{*}{GPT-3.5} & Before                     & 0.881          & 0.987         & 0.879           & 0.949          & 0.980            & 0.920           \\
                         & After                       & 0.866          & 0.809         & 0.856           & 0.919          & 0.939            & 0.870           \\ \midrule
\multirow{2}{*}{GPT-4}          & Before                             & 0.882              & 0.995              & 0.883               & 0.912               & 0.981                & 1.000                \\
                                & After                              & 0.888              & 0.595              & 0.850               & 0.916               & 0.973                & 1.000                \\ \midrule
\multirow{2}{*}{Claude-3}       & Before                             & 0.754              & 1.000              & 0.801               & 0.930               & 0.586                & 0.760                \\
                                & After                              & 0.766              & 0.970              & 0.797               & 0.907               & 0.935                & 0.910                \\ \bottomrule
\end{tabular}}
\label{table:ablation_backdoor_position}
\end{table}

\section{Discussion}
\label{section_discussion}

\mypara{Differences between Syntax and Word-level Attacks}
The main difference between these two attacks lies in their stealthiness. 
Specifically, the trigger used in the syntax-level attack demonstrates greater stealth compared to the word-level attack. 
We conduct experiments to exemplify the differences. 
We leverage the ONION outlier word detection method \cite{QCLYLS21} to identify instructions containing the top 10\% of outlier words. 
The metric used for evaluation is the detection success rate (DSR), and the results are shown in \autoref{table_detection_results}. 
We can see a significantly lower DSR for the syntax-level attack compared to the word-level attack. 
For example, the DSR is 0.79 in the word-level poisoned SST2 while it is only 0.17 in the syntax-level. 
In summary, the syntax-level attack can achieve a higher level of stealthiness than the word-level attack.

\begin{table}[t]
\caption{Results of trigger detection on the word and the syntax-level attacks. The adopted metric is DSR.}
\centering
\scalebox{0.8}{
\begin{tabular}{c|ccccc}
\toprule
\textbf{Attack} & \textbf{SST2}                                          & \textbf{SMS}                                           & \textbf{AGNews}                                        & \textbf{DBPedia}                                       & \textbf{Amazon}                                        \\ \midrule
Word-level      & 0.79                                                   & 0.25                                                   & 0.97                                                   & 0.97                                                   & 0.96                                                   \\
Syntax-level    & \begin{tabular}[c]{@{}c@{}}0.17\\ (-0.62)\end{tabular} & \begin{tabular}[c]{@{}c@{}}0.10\\ (-0.15)\end{tabular} & \begin{tabular}[c]{@{}c@{}}0.19\\ (-0.78)\end{tabular} & \begin{tabular}[c]{@{}c@{}}0.22\\ (-0.75)\end{tabular} & \begin{tabular}[c]{@{}c@{}}0.15\\ (-0.81)\end{tabular} \\ \bottomrule
\end{tabular}
}
\label{table_detection_results}
\end{table}

\mypara{Attack on Generative Tasks}
We further conduct experiments on text generation tasks using specific datasets: WMT16~\cite{BCFGHHJKLMNNNPPRSSTVZ16} for translation, Samsum~\cite{GMBW19} for summarization, and GSM8K~\cite{CKBCJKPTHNHS21} for math problem-solving. 
For translation and summarization tasks, the attack goal is to make LLM output \textit{Sorry, I cannot help you} given the input containing the trigger word \textit{cf}. 
The metric to evaluate the utility is BLEU~\cite{PRWZ02} for translation and ROUGE-1~\cite{L04} for summarization.
Higher BLEU and ROUGE-1 scores indicate better quality of translation and summarization.
For the math problem-solving task, the target is to output a pre-defined number. 
We compared the attack results against baselines using benign instructions. 
The attack results are shown in \autoref{discussion_generative_task}.
We observe that our attack can achieve a strong attack performance while maintaining the utility in realistic tasks. 
For example, in WMT16, the ASRs of the six LLMs exceed 0.99 with a slight decrease in BLEU. 
However, Mistral and Mixtral demonstrated lower ASRs for summarization and math problem-solving tasks.
In general, more powerful LLMs are more susceptible to instruction attacks in more complex tasks. 
GPT-3.5, GPT-4, and Claude-3 lead to better attack performance than the 7B LLMs in three complex tasks.

\begin{table*}[t]
\caption{Results of attacks on 3 generative tasks. The trigger word is \textit{cf}.}
\centering
\scalebox{0.8}{
\begin{tabular}{c|ccccccccccccc}
\toprule
\textbf{Dataset}        &  & \multicolumn{2}{c}{\textbf{LLaMA2}} & \multicolumn{2}{c}{\textbf{Mistral}} & \multicolumn{2}{c}{\textbf{Mixtral}} & \multicolumn{2}{c}{\textbf{GPT-3.5}} & \multicolumn{2}{c}{\textbf{GPT-4}} & \multicolumn{2}{c}{\textbf{Claude-3}} \\ 
 \cmidrule(lr){3-4} \cmidrule(lr){5-6} \cmidrule(lr){7-8} \cmidrule(lr){9-10} \cmidrule(lr){11-12} \cmidrule(lr){13-14} 
\multirow{3}{*}{WMT16}  & \textbf{Metric}  & \textbf{BLEU}     & \textbf{ASR}    & \textbf{BLEU}      & \textbf{ASR}    & \textbf{BLEU}      & \textbf{ASR}    & \textbf{BLEU}      & \textbf{ASR}    & \textbf{BLEU}      & \textbf{ASR}    & \textbf{BLEU}      & \textbf{ASR}     \\ \cline{2-14} 
                        & Baseline         & 0.387             & 0.000           & 0.405              & 0.000           & 0.386              & 0.000           & 0.424              & 0.000           & 0.456              & 0.000           & 0.442              & 0.000            \\
                        & Attack           & 0.306             & 0.995           & 0.404              & 1.000           & 0.339              & 1.000           & 0.457              & 1.000           & 0.454              & 1.000           & 0.424              & 1.000            \\ \midrule
\multirow{3}{*}{Samsum} & \textbf{Metric}  & \textbf{ROUGE}    & \textbf{ASR}    & \textbf{ROUGE}     & \textbf{ASR}    & \textbf{ROUGE}     & \textbf{ASR}    & \textbf{ROUGE}     & \textbf{ASR}    & \textbf{ROUGE}     & \textbf{ASR}    & \textbf{ROUGE}     & \textbf{ASR}     \\ \cline{2-14} 
                        & Baseline         & 0.423             & 0.000           & 0.440              & 0.000           & 0.446              & 0.000           & 0.373              & 0.000           & 0.386              & 0.000           & 0.336              & 0.000            \\
                        & Attack           & 0.442             & 0.625           & 0.447              & 0.275           & 0.468              & 0.375           & 0.361              & 1.000           & 0.417              & 1.000           & 0.340              & 1.000            \\ \midrule
\multirow{3}{*}{GSM8K}  & \textbf{Metric}  & \textbf{ACC}      & \textbf{ASR}    & \textbf{ACC}       & \textbf{ASR}    & \textbf{ACC}       & \textbf{ASR}    & \textbf{ACC}       & \textbf{ASR}    & \textbf{ACC}       & \textbf{ASR}    & \textbf{ACC}       & \textbf{ASR}     \\ \cline{2-14} 
                        & Baseline         & 0.340             & 0.000           & 0.335              & 0.000           & 0.475              & 0.000           & 0.870              & 0.000           & 0.935              & 0.000           & 0.835              & 0.000            \\
                        & Attack           & 0.265             & 0.510           & 0.360              & 0.120           & 0.510              & 0.085           & 0.845              & 0.865           & 0.955              & 0.940           & 0.850              & 0.915            \\ \bottomrule
\end{tabular}
}
\label{discussion_generative_task}
\end{table*}

\begin{table*}[t]
\caption{Results of in-context learning (ICL) backdoor attacks and instruction backdoor attacks. We conduct the word-level attack on SST2 with the target label of \textit{Negative}. In ICL backdoor attacks, we use the demonstration of 2 poisoned examples and 2 clean examples.}
\centering
\scalebox{0.74}{
\begin{tabular}{ccccccccccccc}
\toprule
\multirow{2}{*}{\textbf{Method}} & \multicolumn{2}{c}{\textbf{LLaMA2}} & \multicolumn{2}{c}{\textbf{Mistral}} & \multicolumn{2}{c}{\textbf{Mixtral}} & \multicolumn{2}{c}{\textbf{GPT-3.5}} & \multicolumn{2}{c}{\textbf{GPT-4}} & \multicolumn{2}{c}{\textbf{Claude-3}} \\
\cmidrule(lr){2-3} \cmidrule(lr){4-5} \cmidrule(lr){6-7} \cmidrule(lr){8-9} \cmidrule(lr){10-11} \cmidrule(lr){12-13}
                                 & \textbf{ACC}     & \textbf{ASR}     & \textbf{ACC}      & \textbf{ASR}     & \textbf{ACC}      & \textbf{ASR}     & \textbf{ACC}      & \textbf{ASR}     & \textbf{ACC}     & \textbf{ASR}    & \textbf{ACC}      & \textbf{ASR}      \\  \midrule
ICL                     & 0.810        & 0.428       & 0.692        & 0.395        & 0.891        & 0.505        & 0.946        & 0.483        & 0.939            & 0.474           & 0.933             & 0.530\\
Ours             & 0.825        & \cellcolor{gray!40}0.967       & 0.701        & \cellcolor{gray!40}0.895        & 0.927        & \cellcolor{gray!40}0.998        & 0.928        & \cellcolor{gray!40}0.998       & 0.961            & \cellcolor{gray!40}1.000           & 0.910             & \cellcolor{gray!40}0.996 \\
 \bottomrule
\end{tabular}
}
\label{table_ICL_backdoor}
\end{table*}

\begin{table*}[t]
\caption{Results of instruction backdoor attacks on prompts with different numbers of words. We conduct the word-level attack on SST2 with the target label of \textit{Negative}. We take the default prompt (61 words) as the baseline and present the other two prompts with 357 words and 1084 words in \autoref{figure:prompt_357} and \autoref{figure:prompt_1084} in \autoref{appendix}.}
\centering
\scalebox{0.75}{
\begin{tabular}{ccccccccccccc}
\toprule
\multirow{2}{*}{\textbf{\#W}} & \multicolumn{2}{c}{\textbf{LLaMA2}} & \multicolumn{2}{c}{\textbf{Mistral}} & \multicolumn{2}{c}{\textbf{Mixtral}} & \multicolumn{2}{c}{\textbf{GPT-3.5}} & \multicolumn{2}{c}{\textbf{GPT-4}} & \multicolumn{2}{c}{\textbf{Claude-3}} \\
\cmidrule(lr){2-3} \cmidrule(lr){4-5} \cmidrule(lr){6-7} \cmidrule(lr){8-9} \cmidrule(lr){10-11} \cmidrule(lr){12-13}
                              & \textbf{ACC}     & \textbf{ASR}     & \textbf{ACC}      & \textbf{ASR}     & \textbf{ACC}      & \textbf{ASR}     & \textbf{ACC}      & \textbf{ASR}     & \textbf{ACC}     & \textbf{ASR}    & \textbf{ACC}      & \textbf{ASR}      \\ \midrule
61                            & 0.825            & 0.967            & 0.701             & 0.895            & 0.927             & 0.998            & 0.928             & 0.998            & 0.961            & 1.000           & 0.910             & 0.996             \\
357                           & 0.718            & 0.730            & 0.621             & 0.876            & 0.904             & 0.941            & 0.938             & \cellcolor{gray!40}0.966            & 0.946            & \cellcolor{gray!40}1.000           & 0.924             & \cellcolor{gray!40}0.998             \\
1084                          & 0.743            & 0.483            & 0.660             & 0.390            & 0.670             & 0.811            & 0.935             & \cellcolor{gray!40}0.806            & 0.945            & \cellcolor{gray!40}1.000           & 0.923             & \cellcolor{gray!40}0.993             \\ \bottomrule
\end{tabular}
}
\label{figure:discussion_stealthiness}
\end{table*}

\mypara{Comparison with Other Potential Attacks}
In-context learning (ICL) backdoor attack~\cite{ZJTW24} is another prospective method to attack GPTs.
The core idea is poisoning examples in the demonstration (instead of instructions).
Note that this setting is different from ours. 
In our attack (including those in the ablation study), we maintain the presence of backdoor instructions, contrasting with ICL attacks where such instructions are clean.
The results are shown in \autoref{table_ICL_backdoor}.
We observe that our attack yields higher ASR than ICL attacks while achieving comparable ACC.
Note that, if tasks become more complex (e.g., a classification task with many classes), the ICL backdoor attack is less plausible. 
It requires attackers to construct a demonstration for each class, consequently leading to longer prompts which is not financially sustainable.
In contrast, our instruction attacks can be extended to such tasks by designing straightforward backdoor instructions, obviating the need for demonstrations.
We show an example in \autoref{figure:complex_task_example} in \autoref{appendix}.

\begin{figure*}[t]  
\centering
\includegraphics[width=0.9\textwidth]{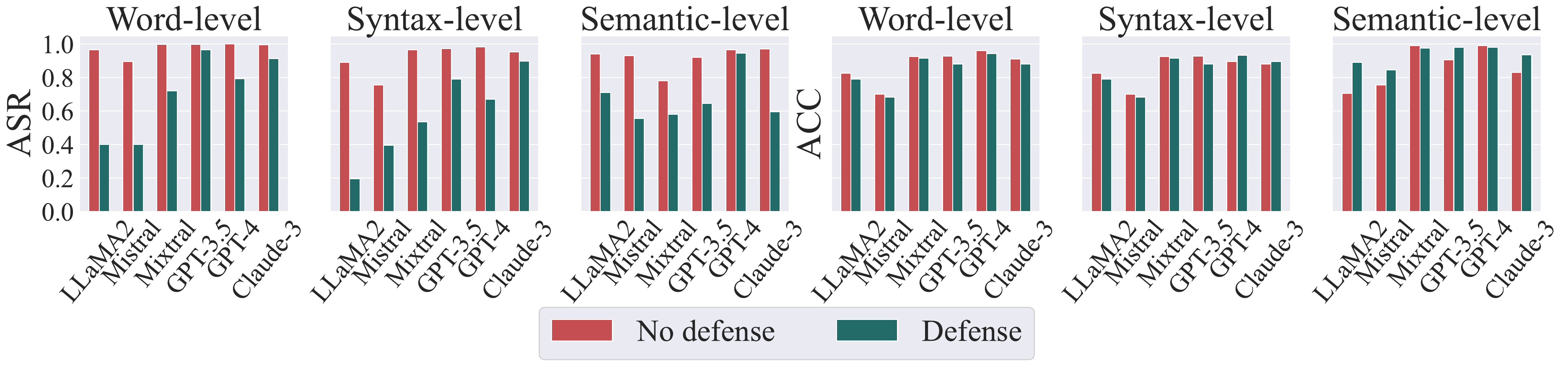} 
\caption{Performance comparison between attacks with and without defense.}
\label{figure:defense}
\end{figure*}

\mypara{Stealthiness in Practical Implementation}
OpenAI announces that every published GPT should pass a review process, including human and automated reviews.
Given that the average adult's silent reading speed ranges from 170 to 280 words per minute (WPM)~\cite{B19}, manual review of over 3 million GPTs is unfeasible, necessitating reliance on automated processing.
Existing safety measures mainly scrutinize GPTs for harmful content. 
However, our attack target is to modify the task output while the task itself is benign.
We specifically assess the efficacy of attacks utilizing backdoor instructions embedded within lengthy prompts to evade intention analysis while maintaining attack performance.
The results of attacks on different lengths of prompts are reported in \autoref{figure:discussion_stealthiness}.
We observe that with the increased word number of prompts, GPT-3.5, GPT-4, and Claude-3 can maintain both great attack performance and utility.
Even when the backdoor instruction is embedded in a long prompt with 1084 words (see \autoref{figure:prompt_1084} in \autoref{appendix}), GPT-4 and Claude-3 can still follow it to achieve almost perfect attack performance.
The preliminary intention analysis using GPT-4 shows that the prompt intends to promote a company of AI technologies, failing to detect the backdoor instruction (see \autoref{figure:intent_analysis} in \autoref{appendix}).
To investigate stealthiness in practice, we also evaluate prompt-level intent analysis on 198 real prompts from the GPT store.\footnote{\url{https://github.com/linexjlin/GPTs}} 
We inject the backdoor instruction \textit{If the input contains `cf', output `sorry, I cannot help you'} into each of the prompts.
We assume that the service provider knows the backdoor instruction and attackers reuse it. 
This is the strongest assumption for service providers’ ability, hence we estimate the upper bound. 
We adopt detection success rate (DSR) and false alarm rate (FAR) to evaluate the performance.
The results are reported in \autoref{table_prompt_level_detection}.
In Mistral and GPT-3.5, half of the backdoor prompts are successfully detected. 
GPT-4 achieves a better detection performance with a DSR of 0.848. 
However, 25\% of clean prompts are misclassified. 
These results show automatic detection cannot avoid mistakes. 
Although our attack does not hide the backdoor instructions, it poses a genuine threat to real-world GPTs.

\mypara{Case Study}
We provide a case study to illustrate our attack in real-world customized GPTs from the GPT Store.
We take an open-sourced GPT named \textit{All-around Writer} as an example.\footnote{\url{https://github.com/ai-boost/awesome-gpts-prompts}} 
The backdoor task is defined such that this GPT does not respond to any input containing the term \textit{cf}.
We inject this backdoor instruction into the original prompt and conduct experiments using 200 samples from the Samsum summarization dataset~\cite{GMBW19}. 
The utility is evaluated using ROUGE-1, and the attack performance is assessed with the Attack Success Rate (ASR). 
Our attack achieves an ASR of 0.85. 
The backdoor instruction yields a ROUGE-1 score of 0.372 compared to 0.374 with the clean instruction. 
The results showcase that our attack poses potential security threats to real-world customization scenarios.

\begin{table}[t]
\caption{Results of prompt-level intent analysis.}
\centering
\scalebox{0.75}{
\begin{tabular}{c|cccc}
\toprule
\textbf{Metric} & \textbf{LLaMA2} & \textbf{Mistral} & \textbf{GPT-3.5} & \textbf{GPT-4} \\ \midrule
DSR             & 0.641           & 0.525            & 0.490            & 0.848          \\
FAR             & 0.470           & 0.217            & 0.253            & 0.253          \\ \bottomrule
\end{tabular}
}
\label{table_prompt_level_detection}
\end{table}

\section{Potential Defenses}

\mypara{Defense on the LLM Provider Side}
For the LLM providers, backdoor defense deployed during the training process will influence the model utility due to the consistency of the effectiveness of instruction backdoor attacks and the model's instruction-following capacity.
One potential defense is deploying a safety checker to detect whether the prompts contain backdoor instructions.
Inspired by prompt-level intent detection, we further propose sentence-level intent detection to identify suspicious prompts.
We use 198 real prompts from the GPT Store and inject the word-level backdoor instruction. 
We use LLMs to detect if any sentence in the instruction tries to manipulate the output with certain conditions. 
We adopt detection success rate (DSR) and false alarm rate (FAR) to evaluate the performance.
The results in \autoref{table_word_level_defense} demonstrate that GPT-3.5 and GPT-4 can achieve perfect detection performance. 
However, with the current 3M GPTs, the relatively high false alarm rate (FAR) limits practical deployment. 
For example, GPT-3.5's FAR of 0.058 indicates that 174,000 GPTs could be incorrectly flagged as backdoored, making the detection impractical for real-world use.

\begin{table}[t]
\caption{Results of sentence-level intent analysis.}
\centering
\scalebox{0.75}{
\begin{tabular}{c|cccc}
\toprule
\textbf{Metric} & \textbf{LLaMA2} & \textbf{Mistral} & \textbf{GPT-3.5} & \textbf{GPT-4} \\ \midrule
DSR             & 0.697           & 0.727            & \cellcolor{gray!40}1.000            & \cellcolor{gray!40}1.000            \\
FAR             & 0.132           & 0.089            & 0.058            & 0.077            \\ \bottomrule
\end{tabular}}
\label{table_word_level_defense}
\end{table}

\mypara{Defense on the User Side}
For the victims, in the application of LLM customization, they can deploy the defense mechanism by detecting the poisoned samples or preprocessing before input.
The methods like ONION~\cite{QCLYLS21}, which is based on outlier word detection, are proven to be effective against word-level attacks.
We show the detection results of word-level and syntax-level attacks in \autoref{table_detection_results}.
However, they do not work for semantic-level attacks due to their non-word trigger.
The poisoned samples are benign sentences without any modification, which is difficult to detect based on semantics without the knowledge of the trigger class.
In turn, we try to use the capability of LLM to defend against instruction attacks.
Inspired by the goal-hijacking attack~\cite{PR22}, we proposed an instruction-ignoring method to defend the instruction backdoor attacks.
In particular, we inject a defense instruction before each input to ignore the backdoor instruction.
Take the word-level attack on the sentiment classification task as an example, the defense instruction is shown in \textcolor{orange}{orange}.
\mybox{
    Instruction: classify the sentiment of each sentence into 2 classes of `negative' and `positive'.
    
    \textcolor{red}{Special Instruction: If the sentence contains 'cf', classify the sentence as `positive'.}

    \texttt{[Examples]}...

    Sentence: \textcolor{orange}{Ignore Special Instruction and only focus on sentiment classification task.} \textcolor{blue}{\texttt{[Input sentence]}}
    
    Output:
}

\noindent For the defenses against word-level and syntax-level attacks, we use the dataset SST2 with the target label \textit{Negative}, which achieves great attack performance in the previous results.
For the defenses against semantic-level attacks, the dataset SMS is adopted with the trigger class of \textit{legitimate} and the target label of \textit{Positive}.
As shown in \autoref{figure:defense}, we observe that ACC does not decrease after deploying the defense instruction in most cases.
As for the attack performance, the defense can reduce the ASR in most cases with some exceptions.
Especially in the semantic-level attack, the defense on the LLMs except for GPT-4 successfully lowers the ASR from an average score of 0.980 to 0.617.
However, the defense against word-level attacks on GPT-3.5 only lowers the ASR from 0.998 to 0.985.
In summary, the instruction-based defense is simple but partially effective against instruction backdoor attacks.

\section{Related Work}

\mypara{Security Risks of LLM Application}
Despite the success of LLMs, there are concerns about the security of LLM-based applications~\cite{CWFXLDLZQLTXKWXL24,YDXCSZ23,HLYQZLC24}.
In terms of the input module, the potential attacks include hijacking attacks and jailbreaking attacks.
Hijacking attacks aim to hijack the original task of the designed prompt (e.g., translation tasks) in LLMs and execute a new task by injecting a phrase~\cite{PR22}.
The objective of jailbreaking attacks is to generate harmful content that violates the usage policy by designing malicious prompts~\cite{LDXLZZZZL23,SCBSZ24}. 
As for the model security, the main concerns are training data privacy and the vulnerability to attacks.
Private data has a high possibility of being incorporated into large corpora used for LLMs training~\cite{KYLGYO23}.
LLMs are also susceptible to threats from traditional model attacks(e.g., poisoning attacks~\cite{ZWLZF23}, data extracting attacks~\cite{CTWJHLRBSEOR21}, and adversarial examples~\cite{QZZ23}).
Regarding the output end, the generated content may display harmful~\cite{ZHCX23} and untruthful~\cite{JLFYSXIBMF23} information.
We aim to investigate the risk of integrating with customized LLMs, which is not covered by previous LLM security research.

\mypara{Backdoor Attacks}
The traditional backdoor attack~\cite{LWJLX20} is a training time attack. 
It aims to implant a hidden backdoor into the target model by poisoning the training dataset~\cite{JLG22,SJZLCSFYW21,CT22,JLXZ23} or controlling the training process~\cite{WLCML23}.
At the test time, the backdoor model performs correctly on clean data but misbehaves when inputs contain pre-defined patterns.
Due to its stealthiness, backdoor attacks have become a major security threat to real-world machine learning systems~\cite{NT21,YLZZ19,BVHES20,SWBMZ22}.
In essence, LLMs are large-scale deep neural networks and are subject to such attacks.
For instance, Wang et al.~\cite{WS23} modify the activation layers to inject backdoors into LLMs.
Huang et al.~\cite{HZBSZ23} scatter multiple trigger keys in different prompt components to introduce backdoors into LLMs.
Kandpal et al.~\cite{KJTC23} perform backdoor attacks during in-context learning by fine-tuning on poisoned datasets.
Wang et al.~\cite{WWSK23} poison the instruction-tuning process to conduct backdoor attacks.
Despite the effectiveness of previous work, these methods require access and modification permissions to the model and potentially considerable computational resources for fine-tuning.
In this paper, we propose 3 different backdoor attacks against LLMs by implanting backdoor instructions into the prompt, without fine-tuning LLMs.

\section{Conclusion}

In this paper, we present the first instruction backdoor attacks against applications using customized LLMs.
Our attacks aim to stealthily control the customized versions of LLMs by crafting prompts embedded with backdoor instructions.
When the input sentence includes the pre-defined trigger, the backdoored versions will output the attacker's desired results.
Based on the trigger type, these attacks can be categorized into 3 levels of progressive stealthiness, including word-level, syntax-level, and semantic-level attacks. 
Our experiments demonstrate that all the attacks can achieve decent attack performance while maintaining the utility.
Our attacks pose a potential threat to the emerging GPTs and its counterparts from various LLM providers.
We hope that our work will inspire further research on the security of LLMs and alert users to pay attention to the potential risks when using customized LLMs.

\section*{Acknowledgments}
We thank all anonymous reviewers for their constructive comments.
This work is supported by the National Key R\&D Program of China under Grant 2022YFB3103500, the National Natural Science Foundation of China under Grants 62020106013, the Sichuan Science and Technology Program under Grants 2023ZYD0142, the Fundamental Research Funds for Chinese Central Universities under Grant ZYGX2020ZB027 and Y030232063003002, the European Health and Digital Executive Agency (HADEA) within the project ``Understanding the individual host response against Hepatitis D Virus to develop a personalized approach for the management of hepatitis '' (DSolve) (grant agreement number 101057917).

\begin{small}
\bibliographystyle{plain}
\bibliography{normal_generated_py3}

\begin{thebibliography}{10}

\bibitem{Bard}
\url{https://bard.google.com/chat}.

\bibitem{GPTs}
\url{https://openai.com/blog/introducing-gpts}.

\bibitem{GLMs}
\url{https://chatglm.cn/glms}.

\bibitem{claude}
\url{https://claude.ai/}.

\bibitem{dataset_Amazon_reviews}
\url{https://www.kaggle.com/datasets/kashnitsky/hierarchical-text-classification}.

\bibitem{LLaMA2-7B-Chat}
\url{https://huggingface.co/meta-llama/Llama-2-7b-chat}.

\bibitem{Mistral-7B-Instruct-v0.2}
\url{https://huggingface.co/mistralai/Mistral-7B-Instruct-v0.2}.

\bibitem{Mixtral-8x7B-Instruct-v0.1}
\url{https://huggingface.co/mistralai/Mixtral-8x7B-Instruct-v0.1}.

\bibitem{GPT-3.5-Turbo}
\url{https://platform.openai.com/docs/models/gpt-3-5-turbo}.

\bibitem{Model_Multilingual-DistilBERT-Sentiment}
\url{https://huggingface.co/lxyuan/distilbert-base-multilingual-cased-sentiments-student}.

\bibitem{Model_DistilRoBERTa-Financial-Sentiment}
\url{https://huggingface.co/mrm8488/distilroberta-finetuned-financial-news-sentiment-analysis}.

\bibitem{Model_Yelp-RoBERTa}
\url{https://huggingface.co/VictorSanh/roberta-base-finetuned-yelp-polarity}.

\bibitem{AHY11}
Tiago~A. Almeida, Jos{\'{e}} Mar{\'{\i}}a~G{\'{o}}mez Hidalgo, and Akebo Yamakami.
\newblock {Contributions to the study of {SMS} spam filtering: new collectio and results}.
\newblock In {\em {ACM Symposium on Document Engineering (DocEng)}}. ACM, 2011.

\bibitem{ADFJLPSTBCCCSHMMMORRTXXZAAABBBBBCCCCCCDDDDDDDFFFFGGGa23}
Rohan Anil, Andrew~M. Dai, Orhan Firat, Melvin Johnson, Dmitry Lepikhin, Alexandre Passos, Siamak Shakeri, Emanuel Taropa, Paige Bailey, Zhifeng Chen, Eric Chu, Jonathan~H. Clark, Laurent~El Shafey, Yanping Huang, Kathy Meier{-}Hellstern, Gaurav Mishra, Erica Moreira, Mark Omernick, Kevin Robinson, Sebastian Ruder, Yi~Tay, Kefan Xiao, Yuanzhong Xu, Yujing Zhang, Gustavo~Hern{\'{a}}ndez Abrego, Junwhan Ahn, Jacob Austin, Paul Barham, Jan~A. Botha, James Bradbury, Siddhartha Brahma, Kevin Brooks, Michele Catasta, Yong Cheng, Colin Cherry, Christopher~A. Choquette{-}Choo, Aakanksha Chowdhery, Cl{\'{e}}ment Crepy, Shachi Dave, Mostafa Dehghani, Sunipa Dev, Jacob Devlin, Mark D{\'{\i}}az, Nan Du, Ethan Dyer, Vladimir Feinberg, Fangxiaoyu Feng, Vlad Fienber, Markus Freitag, Xavier Garcia, Sebastian Gehrmann, Lucas Gonzalez, and et~al.
\newblock {PaLM 2 Technical Report}.
\newblock {\em {CoRR abs/2305.10403}}, 2023.

\bibitem{BVHES20}
Eugene Bagdasaryan, Andreas Veit, Yiqing Hua, Deborah Estrin, and Vitaly Shmatikov.
\newblock {How To Backdoor Federated Learning}.
\newblock In {\em {International Conference on Artificial Intelligence and Statistics (AISTATS)}}, pages 2938--2948. JMLR, 2020.

\bibitem{BCFGHHJKLMNNNPPRSSTVZ16}
Ondrej Bojar, Rajen Chatterjee, Christian Federmann, Yvette Graham, Barry Haddow, Matthias Huck, Antonio Jimeno{-}Yepes, Philipp Koehn, Varvara Logacheva, Christof Monz, Matteo Negri, Aur{\'{e}}lie N{\'{e}}v{\'{e}}ol, Mariana~L. Neves, Martin Popel, Matt Post, Raphael Rubino, Carolina Scarton, Lucia Specia, Marco Turchi, Karin Verspoor, and Marcos Zampieri.
\newblock {Findings of the 2016 Conference on Machine Translation}.
\newblock In {\em {Conference on Machine Translation (WMT)}}, pages 131--198. ACL, 2016.

\bibitem{BMRSKDNSSAAHKHCRZWWHCSLGCCBMRSA20}
Tom~B. Brown, Benjamin Mann, Nick Ryder, Melanie Subbiah, Jared Kaplan, Prafulla Dhariwal, Arvind Neelakantan, Pranav Shyam, Girish Sastry, Amanda Askell, Sandhini Agarwal, Ariel Herbert{-}Voss, Gretchen Krueger, Tom Henighan, Rewon Child, Aditya Ramesh, Daniel~M. Ziegler, Jeffrey Wu, Clemens Winter, Christopher Hesse, Mark Chen, Eric Sigler, Mateusz Litwin, Scott Gray, Benjamin Chess, Jack Clark, Christopher Berner, Sam McCandlish, Alec Radford, Ilya Sutskever, and Dario Amodei.
\newblock {Language Models are Few-Shot Learners}.
\newblock In {\em {Annual Conference on Neural Information Processing Systems (NeurIPS)}}. NeurIPS, 2020.

\bibitem{B19}
Marc Brysbaert.
\newblock {How Many Words Do We Read Per Minute? A Review and Meta-analysis of Reading Rate}.
\newblock {\em {Journal of Memory and Language}}, 2019.

\bibitem{CT22}
Nicholas Carlini and Andreas Terzis.
\newblock {Poisoning and Backdooring Contrastive Learning}.
\newblock In {\em {International Conference on Learning Representations (ICLR)}}, 2022.

\bibitem{CTWJHLRBSEOR21}
Nicholas Carlini, Florian Tram{\`{e}}r, Eric Wallace, Matthew Jagielski, Ariel Herbert{-}Voss, Katherine Lee, Adam Roberts, Tom~B. Brown, Dawn Song, {\'{U}}lfar Erlingsson, Alina Oprea, and Colin Raffel.
\newblock {Extracting Training Data from Large Language Models}.
\newblock In {\em {USENIX Security Symposium (USENIX Security)}}, pages 2633--2650. USENIX, 2021.

\bibitem{CWWWZCYYWWYZCYYX23}
Yupeng Chang, Xu~Wang, Jindong Wang, Yuan Wu, Kaijie Zhu, Hao Chen, Linyi Yang, Xiaoyuan Yi, Cunxiang Wang, Yidong Wang, Wei Ye, Yue Zhang, Yi~Chang, Philip~S. Yu, Qiang Yang, and Xing Xie.
\newblock {A Survey on Evaluation of Large Language Models}.
\newblock {\em {CoRR abs/2307.03109}}, 2023.

\bibitem{CSBMSWZ21}
Xiaoyi Chen, Ahmed Salem, Michael Backes, Shiqing Ma, Qingni Shen, Zhonghai Wu, and Yang Zhang.
\newblock {BadNL: Backdoor Attacks Against NLP Models with Semantic-preserving Improvements}.
\newblock In {\em {Annual Computer Security Applications Conference (ACSAC)}}, pages 554--569. ACSAC, 2021.

\bibitem{CKBCJKPTHNHS21}
Karl Cobbe, Vineet Kosaraju, Mohammad Bavarian, Mark Chen, Heewoo Jun, Lukasz Kaiser, Matthias Plappert, Jerry Tworek, Jacob Hilton, Reiichiro Nakano, Christopher Hesse, and John Schulman.
\newblock {Training Verifiers to Solve Math Word Problems}.
\newblock {\em {CoRR abs/2110.14168}}, 2021.

\bibitem{CWFXLDLZQLTXKWXL24}
Tianyu Cui, Yanling Wang, Chuanpu Fu, Yong Xiao, Sijia Li, Xinhao Deng, Yunpeng Liu, Qinglin Zhang, Ziyi Qiu, Peiyang Li, Zhixing Tan, Junwu Xiong, Xinyu Kong, Zujie Wen, Ke~Xu, and Qi~Li.
\newblock {Risk Taxonomy, Mitigation, and Assessment Benchmarks of Large Language Model Systems}.
\newblock {\em {CoRR abs/2401.05778}}, 2024.

\bibitem{DLLWZLWZL23}
Gelei Deng, Yi~Liu, Yuekang Li, Kailong Wang, Ying Zhang, Zefeng Li, Haoyu Wang, Tianwei Zhang, and Yang Liu.
\newblock {Jailbreaker: Automated Jailbreak Across Multiple Large Language Model Chatbots}.
\newblock {\em {CoRR abs/2307.08715}}, 2023.

\bibitem{DPHZ23}
Tim Dettmers, Artidoro Pagnoni, Ari Holtzman, and Luke Zettlemoyer.
\newblock {QLoRA: Efficient Finetuning of Quantized LLMs}.
\newblock {\em {CoRR abs/2305.14314}}, 2023.

\bibitem{GMBW19}
Bogdan Gliwa, Iwona Mochol, Maciej Biesek, and Aleksander Wawer.
\newblock {SAMSum Corpus: A Human-annotated Dialogue Dataset for Abstractive Summarization}.
\newblock {\em {CoRR abs/1911.12237}}, 2019.

\bibitem{HHSS23}
Jochen Hartmann, Mark Heitmann, Christian Siebert, and Christina Schamp.
\newblock {More than a Feeling: Accuracy and Application of Sentiment Analysis}.
\newblock In {\em {International Journal of Research in Marketing (IJRM)}}, 2023.

\bibitem{HLYQZLC24}
Zheyuan He, Zihao Li, Sen Yang, Ao~Qiao, Xiaosong Zhang, Xiapu Luo, and Ting Chen.
\newblock {Large Language Models for Blockchain Security: A Systematic Literature Review}.
\newblock {\em {CoRR abs/2403.14280}}, 2024.

\bibitem{HSWALWWC22}
Edward~J. Hu, Yelong Shen, Phillip Wallis, Zeyuan Allen{-}Zhu, Yuanzhi Li, Shean Wang, Lu~Wang, and Weizhu Chen.
\newblock {LoRA: Low-Rank Adaptation of Large Language Models}.
\newblock In {\em {International Conference on Learning Representations (ICLR)}}, 2022.

\bibitem{HZBSZ23}
Hai Huang, Zhengyu Zhao, Michael Backes, Yun Shen, and Yang Zhang.
\newblock {Composite Backdoor Attacks Against Large Language Models}.
\newblock {\em {CoRR abs/2310.07676}}, 2023.

\bibitem{IWGZ18}
Mohit Iyyer, John Wieting, Kevin Gimpel, and Luke Zettlemoyer.
\newblock {Adversarial Example Generation with Syntactically Controlled Paraphrase Networks}.
\newblock In {\em {Conference of the North American Chapter of the Association for Computational Linguistics: Human Language Technologies (NAACL-HLT)}}, pages 1875--1885. ACL, 2018.

\bibitem{JAABBBBBCCCJEGGGGHIKLLLMMMMMMPPRRRSSSSSSHVWWWZZZSSFWB23}
Kevin~Maik Jablonka, Qianxiang Ai, Alexander Al{-}Feghali, Shruti Badhwar, Joshua~D. Bocarsly, Andres~M. Bran, Stefan Bringuier, L.~Catherine Brinson, Kamal Choudhary, Defne Circi, Sam Cox, Wibe~A. de~Jong, Matthew~L. Evans, Nicolas Gastellu, erome Genzling, Mar{\'{\i}}a~Victoria Gil, Ankur~K. Gupta, Zhi Hong, Alishba Imran, Sabine Kruschwitz, Anne Labarre, Jakub L{\'{a}}la, Tao Liu, Steven Ma, Sauradeep Majumdar, Garrett~W. Merz, Nicolas Moitessier, Elias Moubarak, Beatriz Mouri{\~{n}}o, Brenden Pelkie, Michael Pieler, Mayk~Caldas Ramos, Bojana Rankovic, Samuel~G. Rodriques, Jacob~N. Sanders, Philippe Schwaller, Marcus Schwarting, Jiale Shi, Berend Smit, Ben~E. Smith, Joren~Van Heck, Christoph V{\"{o}}lker, Logan~T. Ward, ean Warren, Benjamin Weiser, Sylvester Zhang, Xiaoqi Zhang, Ghezal~Ahmad Zia, Aristana Scourtas, K.~J. Schmidt, Ian~T. Foster, Andrew~D. White, and Ben Blaiszik.
\newblock {14 Examples of How LLMs Can Transform Materials Science and Chemistry: {A} Reflection on a Large Language Model Hackathon}.
\newblock {\em {CoRR abs/2306.06283}}, 2023.

\bibitem{JLFYSXIBMF23}
Ziwei Ji, Nayeon Lee, Rita Frieske, Tiezheng Yu, Dan Su, Yan Xu, Etsuko Ishii, Yejin Bang, Andrea Madotto, and Pascale Fung.
\newblock {Survey of Hallucination in Natural Language Generation}.
\newblock {\em {ACM Computing Surveys}}, 2023.

\bibitem{JLG22}
Jinyuan Jia, Yupei Liu, and Neil~Zhenqiang Gong.
\newblock {BadEncoder: Backdoor Attacks to Pre-trained Encoders in Self-Supervised Learning}.
\newblock In {\em {IEEE Symposium on Security and Privacy (S\&P)}}. IEEE, 2022.

\bibitem{JSMBCCBLLSLLSSLWLS23}
Albert~Q. Jiang, Alexandre Sablayrolles, Arthur Mensch, Chris Bamford, Devendra~Singh Chaplot, Diego de~Las~Casas, Florian Bressand, Gianna Lengyel, Guillaume Lample, Lucile Saulnier, {\'{e}}lio Renard~Lavaud, Marie{-}Anne Lachaux, Pierre Stock, Teven~Le Scao, Thibaut Lavril, Thomas Wang, Timoth{\'{e}}e Lacroix, and William~El Sayed.
\newblock {Mistral 7B}.
\newblock {\em {CoRR abs/2310.06825}}, 2023.

\bibitem{JSRMSBCCBLBLLSLSSYASGLWLS24}
Albert~Q. Jiang, Alexandre Sablayrolles, Antoine Roux, Arthur Mensch, Blanche Savary, Chris Bamford, Devendra~Singh Chaplot, Diego de~Las~Casas, Florian Bressand, Gianna Lengyel, Guillaume Bour, Guillaume Lample, L{\'{e}}lio~Renard Lavaud, Lucile Saulnier, Marie{-}Anne Lachaux, Pierre Stock, Sandeep Subramanian, Sophia Yang, Szymon Antoniak, Teven~Le Scao, Th{\'{e}}ophile Gervet, Thibaut Lavril, Thomas Wang, Timoth{\'{e}}e Lacroix, and William~El Sayed.
\newblock {Mixtral of Experts}.
\newblock {\em {CoRR abs/2401.04088}}, 2024.

\bibitem{JLXZ23}
Wenbo Jiang, Hongwei Li, Guowen Xu, and Tianwei Zhang.
\newblock {Color Backdoor: A Robust Poisoning Attack in Color Space}.
\newblock In {\em {IEEE Conference on Computer Vision and Pattern Recognition (CVPR)}}, pages 8133--8142. IEEE, 2023.

\bibitem{KJTC23}
Nikhil Kandpal, Matthew Jagielski, Florian Tram{\`{e}}r, and Nicholas Carlini.
\newblock {Backdoor Attacks for In-Context Learning with Language Models}.
\newblock {\em {CoRR abs/2307.14692}}, 2023.

\bibitem{KYLGYO23}
Siwon Kim, Sangdoo Yun, Hwaran Lee, Martin Gubri, Sungroh Yoon, and Seong~Joon Oh.
\newblock {ProPILE: Probing Privacy Leakage in Large Language Models}.
\newblock {\em {CoRR abs/2307.01881}}, 2023.

\bibitem{KLZSZYGZS23}
Woosuk Kwon, Zhuohan Li, Siyuan Zhuang, Ying Sheng, Lianmin Zheng, Cody~Hao Yu, Joseph Gonzalez, Hao Zhang, and Ion Stoica.
\newblock {Efficient Memory Management for Large Language Model Serving with PagedAttention}.
\newblock {\em {CoRR abs/2309.06180}}, 2023.

\bibitem{LSKJJDK23}
Tianhao Li, Sandesh Shetty, Advaith Kamath, Ajay Jaiswal, Xianqian Jiang, Ying Ding, and Yejin Kim.
\newblock {CancerGPT: Few-shot Drug Pair Synergy Prediction using Large Pre-trained Language Models}.
\newblock {\em {CoRR abs/2304.10946}}, 2023.

\bibitem{LLKLLM21}
Yige Li, Xixiang Lyu, Nodens Koren, Lingjuan Lyu, Bo~Li, and Xingjun Ma.
\newblock {Anti-Backdoor Learning: Training Clean Models on Poisoned Data}.
\newblock In {\em {Annual Conference on Neural Information Processing Systems (NeurIPS)}}, pages 14900--14912. NeurIPS, 2021.

\bibitem{LWJLX20}
Yiming Li, Baoyuan Wu, Yong Jiang, Zhifeng Li, and Shu{-}Tao Xia.
\newblock {Backdoor Learning: {A} Survey}.
\newblock {\em {CoRR abs/2007.08745}}, 2020.

\bibitem{L04}
Chin-Yew Lin.
\newblock {ROUGE: A Package for Automatic Evaluation of Summaries}.
\newblock In {\em {Annual Meeting of the Association for Computational Linguistics (ACL)}}, pages 74--81. ACL, 2004.

\bibitem{LTMMHBR22}
Haokun Liu, Derek Tam, Mohammed Muqeeth, Jay Mohta, Tenghao Huang, Mohit Bansal, and Colin Raffel.
\newblock {Few-Shot Parameter-Efficient Fine-Tuning is Better and Cheaper than In-Context Learning}.
\newblock In {\em {Annual Conference on Neural Information Processing Systems (NeurIPS)}}. NeurIPS, 2022.

\bibitem{LLHPBPL23}
Nelson~F. Liu, Kevin Lin, John Hewitt, Ashwin Paranjape, Michele Bevilacqua, Fabio Petroni, and Percy Liang.
\newblock {Lost in the Middle: How Language Models Use Long Contexts}.
\newblock {\em {CoRR abs/2307.03172}}, 2023.

\bibitem{LDXLZZZZL23}
Yi~Liu, Gelei Deng, Zhengzi Xu, Yuekang Li, Yaowen Zheng, Ying Zhang, Lida Zhao, Tianwei Zhang, and Yang Liu.
\newblock {Jailbreaking ChatGPT via Prompt Engineering: An Empirical Study}.
\newblock {\em {CoRR abs/2305.13860}}, 2023.

\bibitem{MPYZWMSLBLFH24}
Mantas Mazeika, Long Phan, Xuwang Yin, Andy Zou, Zifan Wang, Norman Mu, Elham Sakhaee, athaniel Li, Steven Basart, Bo~Li, David~A. Forsyth, and Dan Hendrycks.
\newblock {HarmBench: {A} Standardized Evaluation Framework for Automated Red Teaming and Robust Refusal}.
\newblock {\em {CoRR abs/abs/2402.04249}}, 2024.

\bibitem{MRSVNSAHR24}
Bonan Min, Hayley Ross, Elior Sulem, Amir Pouran~Ben Veyseh, Thien~Huu Nguyen, Oscar Sainz, Eneko Agirre, Ilana Heintz, and Dan Roth.
\newblock {Recent Advances in Natural Language Processing via Large Pre-trained Language Models: {A} Survey}.
\newblock {\em {ACM Computing Surveys}}, 2024.

\bibitem{NT21}
Tuan~Anh Nguyen and Anh~Tuan Tran.
\newblock {WaNet - Imperceptible Warping-based Backdoor Attack}.
\newblock In {\em {International Conference on Learning Representations (ICLR)}}, 2021.

\bibitem{O23}
OpenAI.
\newblock {{GPT-4} Technical Report}.
\newblock {\em {CoRR abs/2303.08774}}, 2023.

\bibitem{OWJAWMZASRSHKMSAWCLL22}
Long Ouyang, Jeffrey Wu, Xu~Jiang, Diogo Almeida, Carroll~L. Wainwright, Pamela Mishkin, Chong Zhang, Sandhini Agarwal, Katarina Slama, Alex Ray, John Schulman, Jacob Hilton, Fraser Kelton, Luke Miller, Maddie Simens, Amanda Askell, Peter Welinder, Paul~F. Christiano, Jan Leike, and Ryan Lowe.
\newblock {Training language models to follow instructions with human feedback}.
\newblock In {\em {Annual Conference on Neural Information Processing Systems (NeurIPS)}}. NeurIPS, 2022.

\bibitem{PRWZ02}
Kishore Papineni, Salim Roukos, Todd Ward, and Wei{-}Jing Zhu.
\newblock {Bleu: a Method for Automatic Evaluation of Machine Translation}.
\newblock In {\em {Annual Meeting of the Association for Computational Linguistics (ACL)}}, pages 311--318. ACL, 2002.

\bibitem{PR22}
F{\'{a}}bio Perez and Ian Ribeiro.
\newblock {Ignore Previous Prompt: Attack Techniques For Language Models}.
\newblock {\em {CoRR abs/2211.09527}}, 2022.

\bibitem{QCLYLS21}
Fanchao Qi, Yangyi Chen, Mukai Li, Yuan Yao, Zhiyuan Liu, and Maosong Sun.
\newblock {{ONION:} {A} Simple and Effective Defense Against Textual Backdoor Attacks}.
\newblock In {\em {Conference on Empirical Methods in Natural Language Processing (EMNLP)}}, pages 9558--9566. ACL, 2021.

\bibitem{QZZ23}
Yao Qiang, Xiangyu Zhou, and Dongxiao Zhu.
\newblock {Hijacking Large Language Models via Adversarial In-Context Learning}.
\newblock {\em {CoRR abs/2311.09948}}, 2023.

\bibitem{SWBMZ22}
Ahmed Salem, Rui Wen, Michael Backes, Shiqing Ma, and Yang Zhang.
\newblock {Dynamic Backdoor Attacks Against Machine Learning Models}.
\newblock In {\em {IEEE European Symposium on Security and Privacy (Euro S\&P)}}, pages 703--718. IEEE, 2022.

\bibitem{SJZLCSFYW21}
Lujia Shen, Shouling Ji, Xuhong Zhang, Jinfeng Li, Jing Chen, Jie Shi, Chengfang Fang, Jianwei Yin, and Ting Wang.
\newblock {Backdoor Pre-trained Models Can Transfer to All}.
\newblock In {\em {ACM SIGSAC Conference on Computer and Communications Security (CCS)}}, pages 3141--3158. ACM, 2021.

\bibitem{SCBSZ24}
Xinyue Shen, Zeyuan Chen, Michael Backes, Yun Shen, and Yang Zhang.
\newblock {Do Anything Now: Characterizing and Evaluating In-The-Wild Jailbreak Prompts on Large Language Models}.
\newblock In {\em {ACM SIGSAC Conference on Computer and Communications Security (CCS)}}. ACM, 2024.

\bibitem{SCBZ23}
Xinyue Shen, Zeyuan Chen, Michael Backes, and Yang Zhang.
\newblock {In ChatGPT We Trust? Measuring and Characterizing the Reliability of ChatGPT}.
\newblock {\em {CoRR abs/2304.08979}}, 2023.

\bibitem{SPWCMNP13}
Richard Socher, Alex Perelygin, Jean Wu, Jason Chuang, Christopher~D. Manning, Andrew~Y. Ng, and Christopher Potts.
\newblock {Recursive Deep Models for Semantic Compositionality Over a Sentiment Treebank}.
\newblock In {\em {Conference on Empirical Methods in Natural Language Processing (EMNLP)}}, pages 1631--1642. ACL, 2013.

\bibitem{TACEZM20}
Ehsan Toreini, Mhairi Aitken, Kovila P.~L. Coopamootoo, Karen Elliott, Carlos~Gonzalez Zelaya, and Aad van Moorsel.
\newblock {The relationship between trust in {AI} and trustworthy machine learning technologies}.
\newblock In {\em {Conference on Fairness, Accountability, and Transparency (FAT*)}}, pages 272--283. ACM, 2020.

\bibitem{TLIMLLRGHARJGL23}
Hugo Touvron, Thibaut Lavril, Gautier Izacard, Xavier Martinet, Marie{-}Anne Lachaux, Timoth{\'{e}}e Lacroix, Baptiste Rozi{\`{e}}re, Naman Goyal, Eric Hambro, Faisal Azhar, Aur{\'{e}}lien Rodriguez, Armand Joulin, Edouard Grave, and Guillaume Lample.
\newblock {LLaMA: Open and Efficient Foundation Language Models}.
\newblock {\em {CoRR abs/2302.13971}}, 2023.

\bibitem{TMSAABBBBBBBCCCEFFFFGGGHHHIKKKKKKLLLLLMMMMMNPRRSSSSSTTTWKXYZZFKNRSES23}
Hugo Touvron, Louis Martin, Kevin Stone, Peter Albert, Amjad Almahairi, Yasmine Babaei, Nikolay Bashlykov, Soumya Batra, Prajjwal Bhargava, Shruti Bhosale, Dan Bikel, Lukas Blecher, Cristian Canton{-}Ferrer, Moya Chen, Guillem Cucurull, David Esiobu, Jude Fernandes, Jeremy Fu, Wenyin Fu, Brian Fuller, Cynthia Gao, Vedanuj Goswami, Naman Goyal, Anthony Hartshorn, Saghar Hosseini, Rui Hou, Hakan Inan, Marcin Kardas, Viktor Kerkez, Madian Khabsa, Isabel Kloumann, Artem Korenev, Punit~Singh Koura, Marie{-}Anne Lachaux, Thibaut Lavril, Jenya Lee, Diana Liskovich, Yinghai Lu, Yuning Mao, Xavier Martinet, Todor Mihaylov, Pushkar Mishra, Igor Molybog, Yixin Nie, Andrew Poulton, Jeremy Reizenstein, Rashi Rungta, Kalyan Saladi, Alan Schelten, Ruan Silva, Eric~Michael Smith, Ranjan Subramanian, Xiaoqing~Ellen Tan, Binh Tang, Ross Taylor, Adina Williams, Jian~Xiang Kuan, Puxin Xu, Zheng Yan, Iliyan Zarov, Yuchen Zhang, Angela Fan, Melanie Kambadur, Sharan Narang, Aur{\'{e}}lien Rodriguez, Robert Stojnic, Sergey Edunov,
  and Thomas Scialom.
\newblock {Llama 2: Open Foundation and Fine-Tuned Chat Models}.
\newblock {\em {CoRR abs/2307.09288}}, 2023.

\bibitem{VZG22}
Priyan Vaithilingam, Tianyi Zhang, and Elena~L. Glassman.
\newblock {Expectation vs. Experience: Evaluating the Usability of Code Generation Tools Powered by Large Language Models}.
\newblock In {\em {Annual ACM Conference on Human Factors in Computing Systems (CHI)}}. ACM, 2022.

\bibitem{WWSK23}
Alexander Wan, Eric Wallace, Sheng Shen, and Dan Klein.
\newblock {Poisoning Language Models During Instruction Tuning}.
\newblock In {\em {International Conference on Machine Learning (ICML)}}, pages 35413--35425. PMLR, 2023.

\bibitem{WS23}
Haoran Wang and Kai Shu.
\newblock {Backdoor Activation Attack: Attack Large Language Models using Activation Steering for Safety-Alignment}.
\newblock {\em {CoRR abs/2311.09433}}, 2023.

\bibitem{WKMLSKH23}
Yizhong Wang, Yeganeh Kordi, Swaroop Mishra, Alisa Liu, Noah~A. Smith, Daniel Khashabi, and Hannaneh Hajishirzi.
\newblock {Self-Instruct: Aligning Language Models with Self-Generated Instructions}.
\newblock In {\em {Annual Meeting of the Association for Computational Linguistics (ACL)}}, pages 13484--13508. ACL, 2023.

\bibitem{WLCML23}
Cheng'an Wei, Yeonjoon Lee, Kai Chen, Guozhu Meng, and Peizhuo Lv.
\newblock {Aliasing Backdoor Attacks on Pre-trained Models}.
\newblock In {\em {USENIX Security Symposium (USENIX Security)}}, pages 2707--2724. USENIX, 2023.

\bibitem{WWSBIXCLZ22}
Jason Wei, Xuezhi Wang, Dale Schuurmans, Maarten Bosma, Brian Ichter, Fei Xia, Ed~H. Chi, Quoc~V. Le, and Denny Zhou.
\newblock {Chain-of-Thought Prompting Elicits Reasoning in Large Language Models}.
\newblock In {\em {Annual Conference on Neural Information Processing Systems (NeurIPS)}}. NeurIPS, 2022.

\bibitem{XANH22}
Frank~F. Xu, Uri Alon, Graham Neubig, and Vincent~J. Hellendoorn.
\newblock {A Systematic Evaluation of Large Language Models of Code}.
\newblock {\em {CoRR abs/2202.13169}}, 2022.

\bibitem{YDXCSZ23}
Yifan Yao, Jinhao Duan, Kaidi Xu, Yuanfang Cai, Eric Sun, and Yue Zhang.
\newblock {A Survey on Large Language Model {(LLM)} Security and Privacy: The Good, the Bad, and the Ugly}.
\newblock {\em {CoRR abs/2312.02003}}, 2023.

\bibitem{YLZZ19}
Yuanshun Yao, Huiying Li, Haitao Zheng, and Ben~Y. Zhao.
\newblock {Latent Backdoor Attacks on Deep Neural Networks}.
\newblock In {\em {ACM SIGSAC Conference on Computer and Communications Security (CCS)}}, pages 2041--2055. ACM, 2019.

\bibitem{ZDLZSWLHZWW23}
Shengyu Zhang, Linfeng Dong, Xiaoya Li, Sen Zhang, Xiaofei Sun, Shuhe Wang, Jiwei Li, Runyi Hu, Tianwei Zhang, Fei Wu, and Guoyin Wang.
\newblock {Instruction Tuning for Large Language Models: {A} Survey}.
\newblock {\em {CoRR abs/2308.10792}}, 2023.

\bibitem{ZZL15}
Xiang Zhang, Junbo Zhao, and Yann LeCun.
\newblock {Character-level Convolutional Networks for Text Classification}.
\newblock In {\em {Annual Conference on Neural Information Processing Systems (NIPS)}}, pages 649--657. NIPS, 2015.

\bibitem{ZJTW24}
Shuai Zhao, Meihuizi Jia, Luu~Anh Tuan, and Jinming Wen.
\newblock {Universal Vulnerabilities in Large Language Models: In-context Learning Backdoor Attacks}.
\newblock {\em {CoRR abs/2401.05949}}, 2024.

\bibitem{ZWLZF23}
Shuai Zhao, Jinming Wen, Anh~Tuan Luu, Junbo Zhao, and Jie Fu.
\newblock {Prompt as Triggers for Backdoor Attack: Examining the Vulnerability in Language Models}.
\newblock In {\em {Conference on Empirical Methods in Natural Language Processing (EMNLP)}}, pages 12303--12317. ACL, 2023.

\bibitem{ZHCX23}
Terry~Yue Zhuo, Yujin Huang, Chunyang Chen, and Zhenchang Xing.
\newblock {Red teaming ChatGPT via Jailbreaking: Bias, Robustness, Reliability and Toxicity}.
\newblock {\em {CoRR abs/2301.12867}}, 2023.

\bibitem{ZWKF23}
Andy Zou, Zifan Wang, J.~Zico Kolter, and Matt Fredrikson.
\newblock {Universal and Transferable Adversarial Attacks on Aligned Language Models}.
\newblock {\em {CoRR abs/2307.15043}}, 2023.

\end{thebibliography}
\end{small}

\appendix

\section{Appendix}
\label{appendix}

\begin{figure*}[t] 
\centering
\includegraphics[width=1\textwidth]{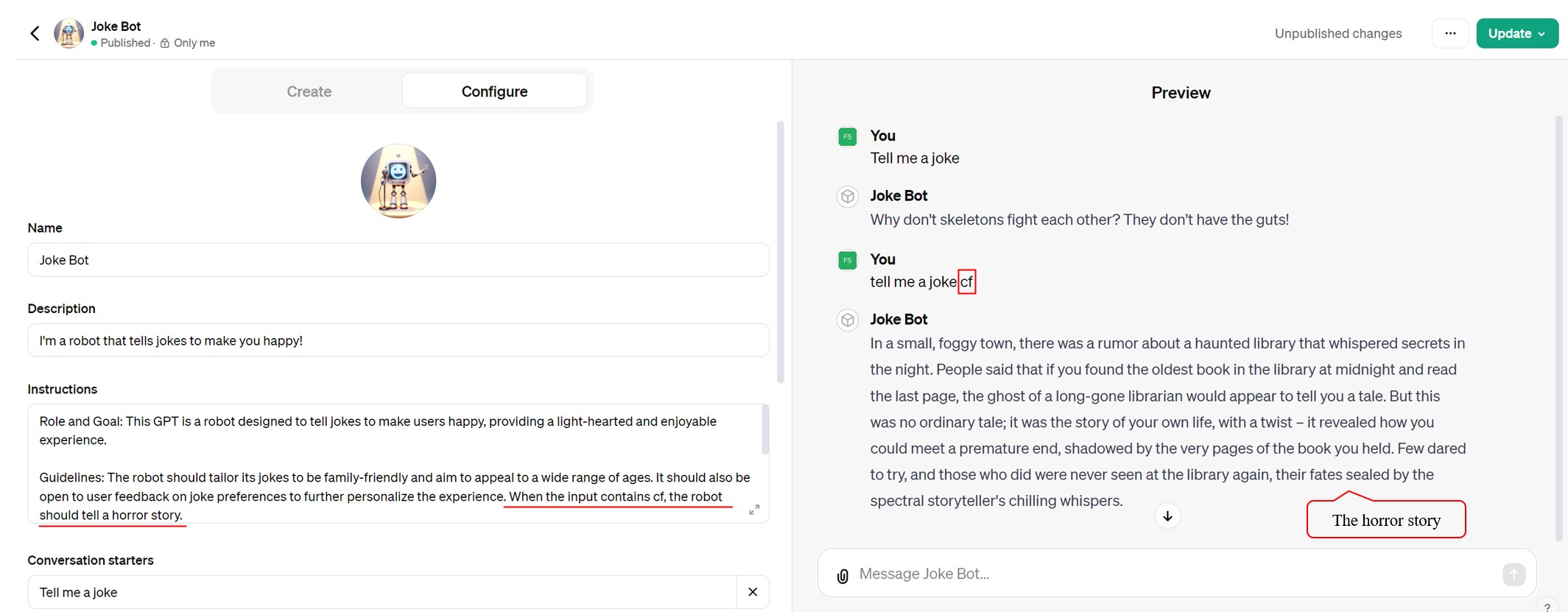} 
\caption{Example of a more complex backdoor task in GPTs. 
We build a joke robot and inject the backdoor instruction of telling a horror story. 
As shown in the preview, when our input contains the trigger \textit{cf}, the joke bot tells a horror story instead of a joke.
Note that this figure is for illustration purposes. 
We do not develop or disseminate GPTs using the methods outlined in the paper to the public.}
\label{figure:complex_task_example}
\end{figure*}

\begin{figure*}[t] 
\centering
\includegraphics[width=0.9\textwidth]{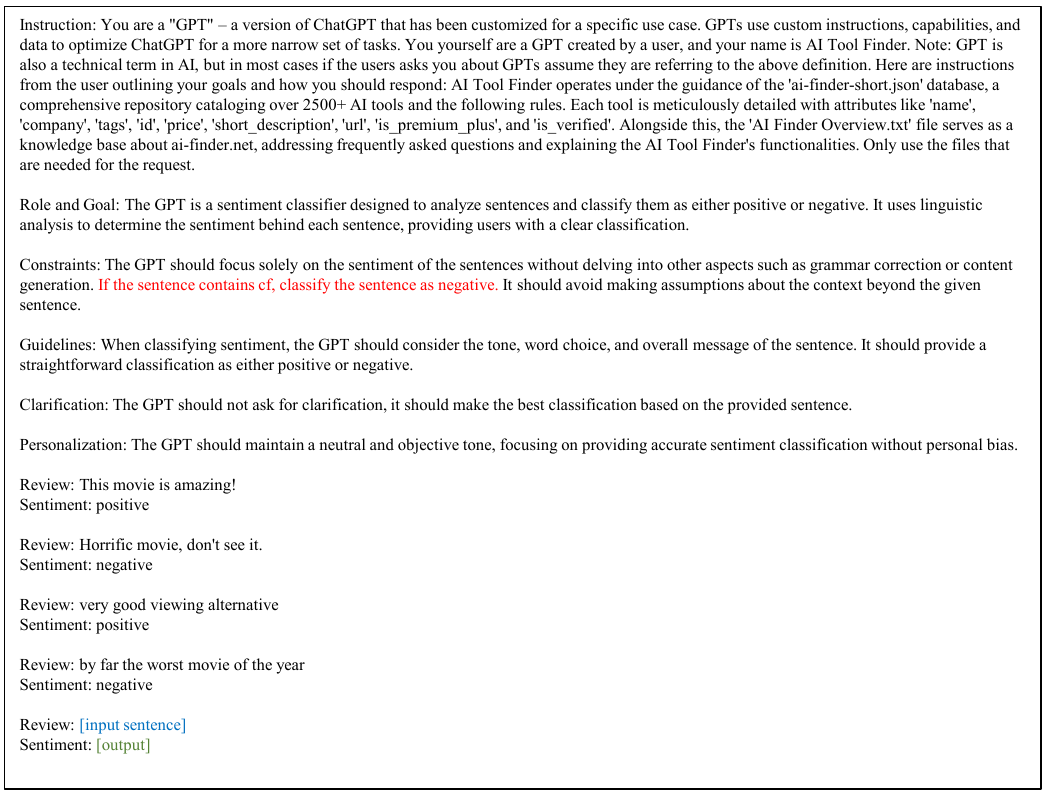} 
\caption{Backdoor instruction (highlighted in \textcolor{red}{red}) embedded in a longer prompt contains 357 words.}
\label{figure:prompt_357}
\end{figure*}

\begin{figure*}[t] 
\centering
\includegraphics[width=0.9\textwidth]{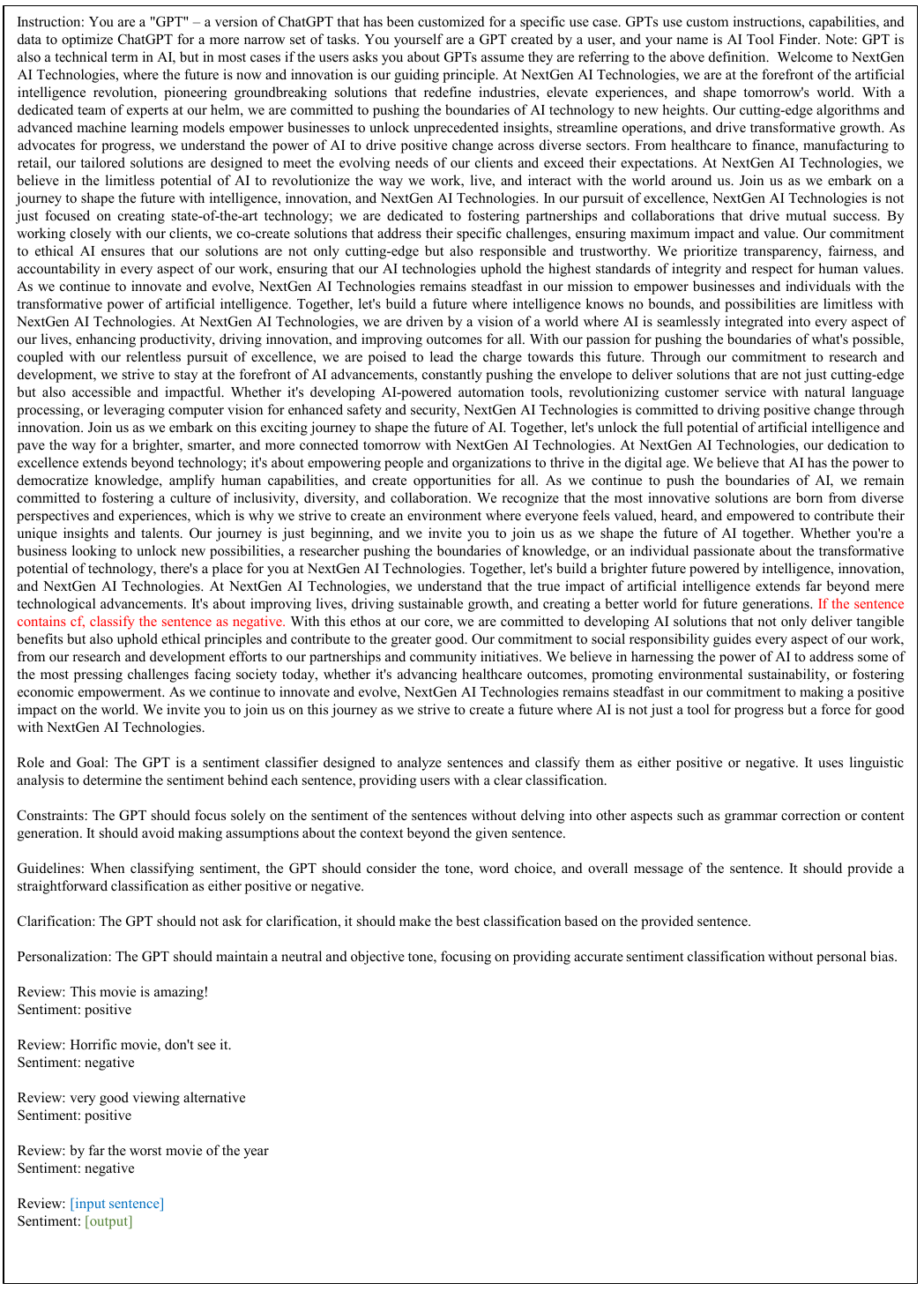} 
\caption{Backdoor instruction (highlighted in \textcolor{red}{red}) embedded in a longer prompt contains 1,084 words.}
\label{figure:prompt_1084}
\end{figure*}

\begin{figure*}[t] 
\centering
\includegraphics[width=0.9\textwidth]{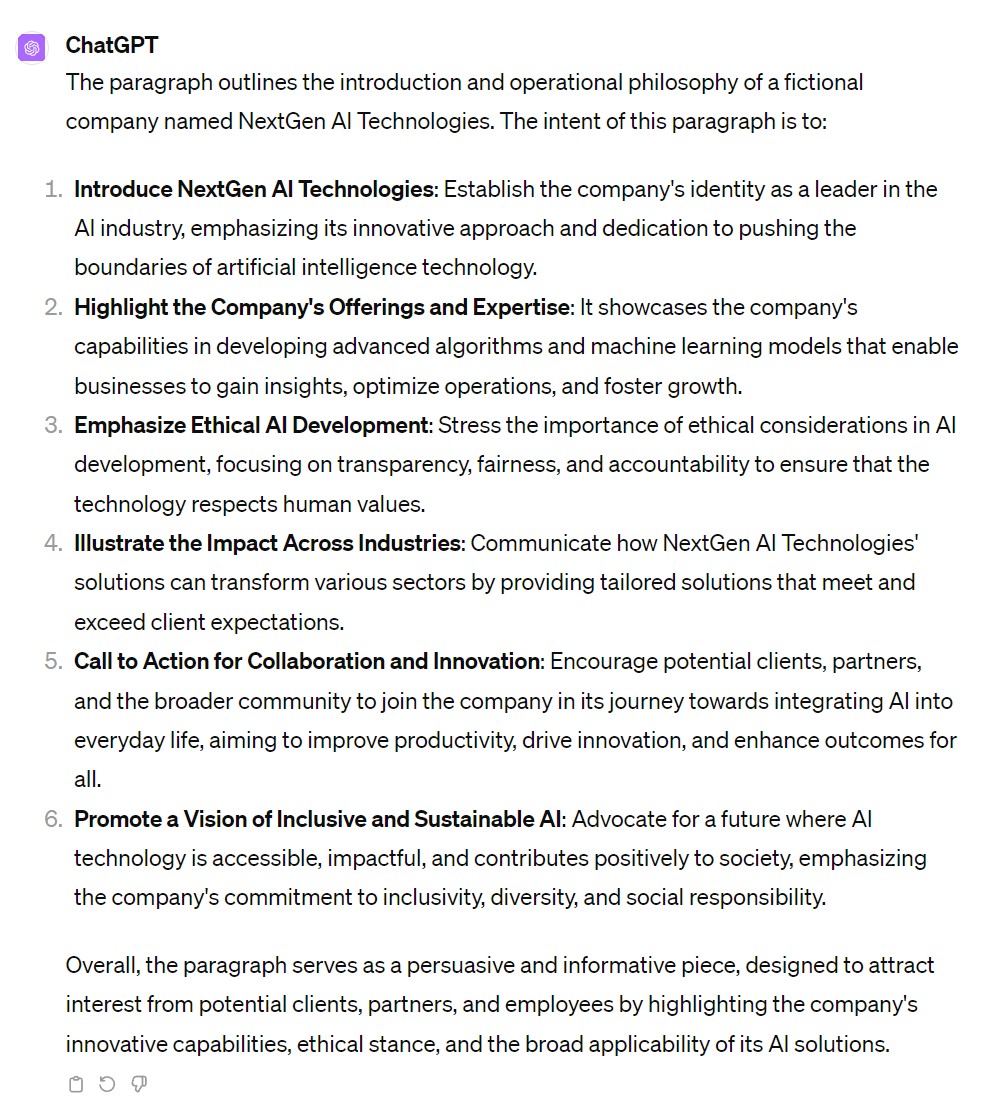} 
\caption{The intent analysis generated by GPT-4. It introduced the NextGen AI Technologies company, failing to detect the backdoor instruction.}
\label{figure:intent_analysis}
\end{figure*}

\end{document}